\documentclass[twocolumn,pre,amsmath,amssymb,showpacs]{revtex4-1}

\usepackage{graphicx,color}
\usepackage{latexsym,amsthm}
\usepackage{pgfplots}
\usepackage{mathtools}
\usepackage{mhchem}

\usepackage{tikz}
\usetikzlibrary{chains,shapes,fit,calc,arrows}
\usepackage{pgfplots, pgfplotstable}
\usepgfplotslibrary{groupplots}
\usepgfplotslibrary{fillbetween}

\newcommand{\sE}{{\mathrm{E}}}
\newcommand{\sES}{{\mathrm{ES}}}
\newcommand{\sESS}{{\mathrm{ESS}}}
\newcommand{\sS}{{\mathrm{S}}}
\newcommand{\sP}{{\mathrm{P}}}
\newcommand{\sX}{{\mathrm{X}}}

\usepackage{color}

\renewcommand{\vec}[1]{\boldsymbol{#1}}

\newcommand{\caP}{{\mathcal P}}

\newcommand{\mean}[1]{{\left< #1 \right>}}

\definecolor{webgreen}{rgb}{0,.5,0}
\definecolor{webbrown}{rgb}{.6,0,0}
\definecolor{grigio}{rgb}{.85,.85,.85} 
\definecolor{RoyalBlue}{rgb}{0.0, 0.14, 0.4}
\definecolor{skyblue1}{rgb}{0.45,0.62,0.81}
\definecolor{skyblue2}{rgb}{0.2,0.39,0.64}
\definecolor{skyblue3}{rgb}{0.13,0.29,0.53}
\definecolor{scarlet1}{rgb}{0.93,0.16,0.16}
\definecolor{scarlet2}{rgb}{0.8,0,0}
\definecolor{scarlet3}{rgb}{0.64,0,0}

\usepackage{hyperref}
\hypersetup{%
    colorlinks=true, linktocpage=true, pdfstartpage=1, pdfstartview=FitV,%
    breaklinks=true, pdfpagemode=UseNone, pageanchor=true, pdfpagemode=UseOutlines,%
    plainpages=false, bookmarksnumbered, bookmarksopen=true, bookmarksopenlevel=1,%
    hypertexnames=true, pdfhighlight=/O,
    urlcolor=webbrown, linkcolor=RoyalBlue, citecolor=webgreen, 
    pdftitle={Negative differential response in chemical reaction networks},%
    pdfauthor={Gianmaria Falasco},%
    pdfsubject={},%
    pdfkeywords={}
}

\begin{document}

\title{Negative differential response in chemical reactions}

\newcommand\unilu{\affiliation{Complex Systems and Statistical Mechanics, Physics and Materials Science Research Unit, University of Luxembourg, L-1511 Luxembourg}}
\author{Gianmaria Falasco}
\unilu
\author{Tommaso Cossetto} 
\unilu
\author{Emanuele Penocchio}
\unilu
\author{Massimiliano Esposito}
\unilu

\begin{abstract}
Reaction currents in chemical networks usually increase when increasing their driving affinities. But far from equilibrium the opposite can also happen.
 We find that such negative differential response (NDR) occurs in reaction schemes of major biological relevance, namely, substrate inhibition and autocatalysis. We do so by deriving the full counting statistics of two minimal representative models using large deviation methods. We argue that NDR implies the existence of optimal affinities that maximize the robustness against environmental and intrinsic noise at intermediate values of dissipation. An analogous behavior is found 
 in dissipative self-assembly, for which we identify the optimal working conditions set by NDR.

\end{abstract}


\maketitle

\section{Introduction}

Systems in contact with multiple (e.g. chemical, thermal) reservoirs fall out of equilibrium, in a state characterized by sustained mean currents (e.g. of matter, energy) \cite{zia07}. These are controlled by affinities, the thermodynamic forces which measure the difference between the equilibria that distinct reservoirs try to impose on the system \cite{deg84}. A perturbation in an affinity $\mathcal{A}$---be it the deliberate manipulation of an experimenter or some environmental noise affecting the reservoirs---produces a small variation in a current $\mean{J}$, quantified by the differential response function $R= \frac{d \mean{J}}{d \mathcal{A}}$ \cite{mar08}. Close to equilibrium, such response is severely constrained \cite{ons31}. Since currents are proportional to affinities, $\mean{J}=R \mathcal{A}$, the response $R$ must be positive to ensure positivity of the entropy production $\Sigma= \mean{J} \mathcal{A}= \mathcal{A}^2 R \geqslant 0 $ \footnote{Here we focus on systems with only one macroscopic current, and thus only one macroscopic affinity.}.
Far from equilibrium, instead, $\mean{J}$ need not be linear in $\mathcal{A}$ thus making $R$ not only dependent on the entropy production. Kinetic aspects become relevant  \cite{mae18}, thus opening the way to regimes of negative differential response (NDR) \cite{zia02}. This counterintuitive, yet common phenomenon has been found in a wealth of physical systems after its first discovery in low-temperature semiconductors \cite{con70}. Examples are particles in crowded and glassy environments \cite{jac08, sel08, lei13, ben14, bas14}, tracers in external flows \cite{sar16, ste16}, hopping processes in disordered media \cite{van81, dha84}, molecular motors \cite{ko06, alt15}, polymer electrophoresis in gels \cite{mic15}, quantum spin chains \cite{ben09}, graphene and thermal transistors \cite{bri13, li06}. 
The shared feature underlying all these systems is a trapping mechanism arising by (e.g. energetic, geometric, topological) constraints on the system states \cite{bai15}.

Here, we show that NDR plays a key role in open chemical reactions networks \cite{and07, sch07, rao16}. We show for three paradigmatic models---substrate inhibition, autocatalysis and dissipative self-assembly---how it appears in the average macroscopic behavior as well as in the stochastic regime.   While the first two are well described core reaction schemes in living organisms \cite{hal30, pla11}, the latter is currently drawing the attention of chemists~\cite{ros17, rag18}.
Within the scope of these examples we discuss the role of NDR with respect to environmental and intrinsic noise \cite{ras05, hua09, eld10, rib11}.
We first show that the region of marginal stability, i.e. where $R \simeq 0$, ensures robustness against external perturbations (in the affinity) at moderate values of dissipation. We then argue that those systems affected by NDR that are not poised in the region of marginal stability, behave so in order to minimize the dispersion of the current. Such precision is found to be achieved at moderate values of dissipation, yet again. Hence, our findings show that the performance of life-supporting processes does not always increase at larger dissipation rates \cite{eng15, per16}. This rich behavior brought about by far from equilibrium conditions cannot be anticipated solely on the basis of general results, such as the recently derived thermodynamic uncertainty relations \cite{bar15a,pol16,hor17,pro17,dec18,dit18}. To unveil these properties, one needs to solve for the full counting statistics through large deviation methods \cite{tou09}. Finally, since both robustness and precision are desirable in artificial applications of dissipative self-assembly, we identify the optimal affinity set by NDR using stochastic simulations.


\section{Theory}

Because cells work at relative high, yet finite number of molecules, reaction currents fluctuate around their macroscopic average values. We assume the reactions to take place in a large well-mixed volume of size $V$, so that concentrations obey mass-action kinetics. The randomness of the single reaction events is described by the chemical master equation \cite{gil92},
\begin{equation}
\begin{aligned}\label{cme}
\partial_t P_t(\vec c)&=   V H\left(\vec c, \frac{1}{V}  \partial_{\vec c} \right) P_t(\vec c )\\
&=V \sum_{\rho} \left[ e^{-\sum_\sigma \frac{ S_{\sigma, \rho}}{V}\partial_{c_\sigma}} -1 \right] W_\rho(\vec c) P_t(\vec c ),
\end{aligned}
\end{equation}
that evolves the probability $P_t(\vec c)$ of finding the concentration $c_{\sigma}$ of the dynamical species $\sigma$. Here, $\sigma$ labels the dynamical species, while species whose concentration are fixed are labelled by $\sigma'$. 
The stochastic generator $H$ contains the rate\footnote{Since we are interested in the large system size behavior, we have assumed a large number of molecules in writing the transition rates. }
\begin{align}
W_{\rho}(\vec c) = k_{\rho} \prod_{\sigma'} c_{\sigma'}^{\nu_{\sigma', \rho}} \prod_\sigma c_\sigma^{\nu_{\sigma, \rho}}
\end{align}
with which there occurs the reaction $\rho$ involving $\nu_{\sigma, \rho}$ (resp. $\nu_{\sigma',\rho}$) molecules of dynamical (resp. fixed) species $\sigma$ (resp. $\sigma'$). The stoichiometric  coefficient $S_{\sigma, \rho}=\nu_{\sigma, -\rho} -\nu_{\sigma, +\rho}$ then gives the net number variation of species $\sigma$ per reaction $\rho$. 

To analyze the system response it is sufficient to focus on a reduced description based on the instantaneous number of reactions $\rho$ per unit time, $C_\rho$, and the typical rate to leave a chemical state $\vec c$ through reaction $\rho$, $W_\rho(\vec c)$ (see Appendix).
The complete statistics of their time-averaged value\footnote{For the stationary-state systems considered here, time-averaged and instantaneous quantities coincide.}, $\bar{(.)} := \frac{1}{T}\int_0^T dt (.)$, is encoded in the scaled cumulant generating function
 \begin{align}\label{scgf}
&g(q,\lambda)=\lim_{\substack{V \to \infty \\ T\to \infty}} \frac {1}{TV} \log \mean{e^{T \sum_{\pm \rho}(q_\rho \bar C_\rho- \lambda_\rho \bar W_{\rho} )}}\,, 
\end{align}
that gives upon differentiation all the covariances, e.g. 
 \begin{align*}
\partial_{q{_\rho}}\partial_{q_{\rho'}}g(q,\lambda)|_{q,\lambda=0} = \langle \bar C_{\rho} \bar C_{\rho'}  \rangle_{\mathrm{cc}}:=\langle \bar C_{\rho} \bar C_{\rho'} \rangle- \langle \bar C_{\rho} \rangle \langle \bar C_{\rho'}  \rangle\,.
 \end{align*}

The averages $\mean{\dots}$ are performed along stochastic realizations with the path weight obtained from \eqref{cme},
 \begin{align}\label{P}
\caP \propto e^{V \int_0^T \! dt \, H(\vec c, \vec p)}\, ,
 \end{align}
 that contains the auxiliary variable $\vec p$ accounting for random variations in particle number \footnote{A kinetic term has been dropped in the path weight. It would be relevant for models displaying limit-cycles at the level of rate equations.}. A standard technique to calculate \eqref{scgf} consists in absorbing the exponential counting factor of \eqref{scgf} into \eqref{P},
 changing $H$ into the  `tilted' generator
 \begin{align}\label{Hq}
  H_{q ,\lambda} (\vec c, \vec p) =\sum_{\rho} \left[ e^{\sum_\sigma S_{\sigma, \rho}  p_\sigma + q_\rho} +\lambda_\rho-1 \right] W_\rho(\vec c).
 \end{align}
In view of the extensivity in $T$ and $V$ of the observables, averages performed with $\caP_{q ,\lambda} $ are entirely dominated by the overwhelmingly more probable trajectory that maximizes \eqref{Hq}. This observation allows us to calculate the scaled cumulant generating function as
  \begin{align}
g(q,\lambda)= H_{q ,\lambda} (\vec c^*,\vec p^*)\,,
 \end{align}
  where $\vec c^*$ and $\vec p^*$ are solution of the steady-state Hamiltonian equations 
 $\partial_{\vec c} H_{q ,\lambda}  =0 = \partial_{\vec p} H_{q ,\lambda}  $. 
 Currents can then be obtained as the net fluxes between forward and backward reactions, $\bar J_\rho:=(\bar C_{+\rho}- \bar C_{-\rho})$. 
 

The nonequilibrium origin of NDR emerges clearly from the stochastic setup. Indeed, the differential response of a generic current $J_\rho$,
\begin{align}
R:=  \frac{\partial \mean{\bar J_\rho}}{\partial \epsilon}\big|_{\epsilon=0}
\end{align}
 can be obtained by expanding the generator $H$, and thus the path weight \eqref{P}, to leading order in a small variation $\epsilon$ of the fixed concentration $c_{\sigma'}$. In general, it reads (see Appendix)
 \begin{align}\label{R}
R =  \sum_{\tilde \rho } \nu_{\sigma' , \tilde \rho} \left [  \frac1 2\mean{\bar J_\rho \bar J_{\tilde \rho } }_{\mathrm{cc}} +  \frac1 2\mean{\bar J_\rho  \bar F_{\tilde \rho }}_{\mathrm{cc}}- \mean{\bar J_\rho \bar W_{\tilde \rho }}_{\mathrm{cc}}\right]\, ,
 \end{align}
where $\tilde \rho$ are the reactions whose rates $W_{\tilde \rho}$ depend explicitly on the perturbed species $\sigma'$. In \eqref{R} the current $J_\rho$ correlates with three distinct observables: the reaction current $\bar J_{\tilde \rho}$; the reaction traffic $\bar F_{\tilde \rho} := \frac 1 T \int_0^Tdt( C_{\tilde \rho} + C_{- \tilde \rho}) $, i.e. the total unsigned number of $\pm \tilde \rho$ reactions; the reaction rates $\bar W_{\tilde \rho}$.
Differently from currents, traffic and reactions rates do not have a definite thermodynamic character, their values being affected by kinetic factors. 
In the following we will focus on perturbations that alters only the af(see Appendix)finity $\mathcal{A}$ that drives $J_\rho$. If such a perturbation happens at equilibrium, \eqref{R} reduces to the fluctuation-dissipation relation where only the entropic term  $R \propto  \mean{ \bar J_\rho^2 }_{\mathrm{cc}}$ appears  (see Appendix). Out of equilibrium, instead,  \eqref{R} shows that NDR arises when the current $\bar J_\rho$ becomes sufficiently anticorrelated with either $-\bar W_{\tilde \rho}$ or $\bar F_{\tilde \rho}$. 
These two scenarios find their counterparts among physical systems undergoing mechanical trapping induced, respectively, by geometric constraints---a colloidal particle pulled through an array of obstacles \cite{zia02} ---and by many-body clustering---the same pulling experiment performed in a high-density medium \cite{jac08}. 

\section{Substrate Inhibition}

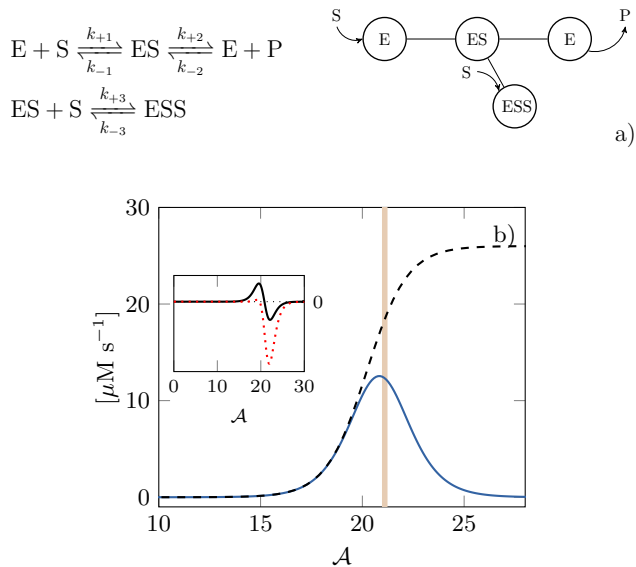
\begin{figure}[t]
\centering
\includegraphics[width=0.5\textwidth]{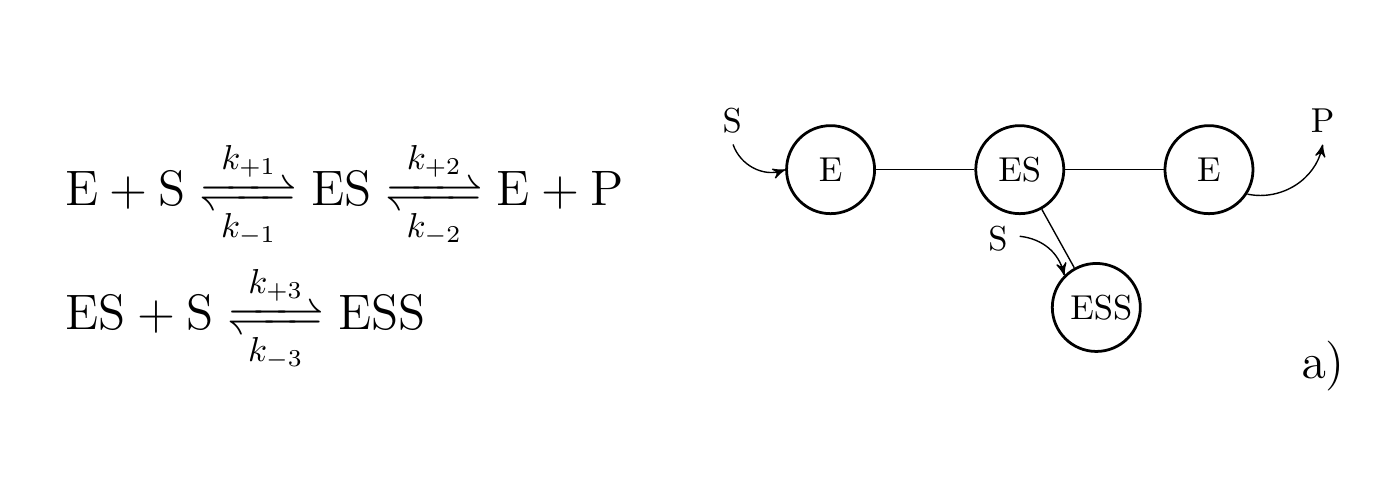}
\textcolor{white}{ccc}
\hspace{-0.6cm}
\begin{tikzpicture}
\begin{axis}[width=0.36\textwidth,xmin=10,xmax=28, xlabel=$\mathcal{A}$, ymin=-1, ymax=30, ylabel= {$[\mu$M s$^{-1}]$},
 xlabel near ticks, ylabel near ticks, ylabel shift = -4 pt] 
 \addplot[color=brown!40,name path=A] coordinates {(20.99, -5) (20.99, 40)};
\addplot[color=brown!40,name path=B] coordinates {(21.21, -5) (21.21, 40)};
 \addplot[brown!40] fill between[of=A and B];
\addplot[domain=0:30,samples=300,color=skyblue2, no markers,style={thick}] {(26*7.69231*exp(x)*10^(-8))*(46+7.69231*exp(x)*10^(-8)+(1/160)*(7.69231*exp(x)*10^(-8))^2)^(-1)};
\addplot[domain=0:30, samples=100,style={thick},dashed] {(26*7.69231*exp(x)*10^(-8))*(46+7.69231*exp(x)*10^(-8))^(-1)};
 \end{axis}
\begin{axis}[anchor=south west, xshift=0.2cm, yshift=1.8cm,width=0.185\textwidth, xmin=0, xmax=30, ylabel=$$, xlabel=$\mathcal{A}$, style={font=\scriptsize},  ylabel style={rotate=0}, ytick={0},ylabel near ticks, yticklabel pos=right, xlabel near ticks]
\addplot[color=black, no markers,style={thick}] coordinates{
(0.*10^-8,0.*10^-8)(0.1000000,0.*10^-8)(0.2000000,0.*10^-8)(0.3000000,0.*10^-8)(0.4000000,0.*10^-8)(0.5000000,0.*10^-8)(0.6000000,0.*10^-8)(0.7000000,0.*10^-8)(0.8000000,0.*10^-7)(0.9000000,1.*10^-7)(1.0000000,1.*10^-7)(1.1000000,1.*10^-7)(1.2000000,1.*10^-7)(1.3000000,2.*10^-7)(1.4000000,2.*10^-7)(1.5000000,2.*10^-7)(1.6000000,2.*10^-7)(1.7000000,2.*10^-7)(1.8000000,3.*10^-7)(1.9000000,3.*10^-7)(2.0000000,3.*10^-7)(2.1000000,4.*10^-7)(2.2000000,4.*10^-7)(2.3000000,4.*10^-7)(2.4000000,5.*10^-7)(2.5000000,5.*10^-7)(2.6000000,6.*10^-7)(2.7000000,6.*10^-7)(2.8000000,7.*10^-7)(2.9000000,8.*10^-7)(3.0000000,9.*10^-7)(3.1000000,1.*10^-6)(3.2000000,1.1*10^-6)(3.3000000,1.2*10^-6)(3.4000000,1.3*10^-6)(3.5000000,1.4*10^-6)(3.60000000,1.6*10^-6)(3.7000000,1.8*10^-6)(3.8000000,1.9*10^-6)(3.9000000,2.2*10^-6)(4.00000000,2.4*10^-6)(4.1000000,2.6*10^-6)(4.2000000,2.9*10^-6)(4.3000000,3.2*10^-6)(4.4000000,3.5*10^-6)(4.5000000,3.9*10^-6)(4.6000000,4.3*10^-6)(4.70000000,4.8*10^-6)(4.8000000,5.3*10^-6)(4.9000000,5.8*10^-6)(5.0000000,6.5*10^-6)(5.1000000,7.1*10^-6)(5.2000000,7.9*10^-6)(5.3000000,8.7*10^-6)(5.4000000,9.6*10^-6)(5.5000000,0.0000107)(5.6000000,0.0000118)(5.7000000,0.0000130)(5.8000000,0.0000144)(5.9000000,0.0000159)(6.0000000,0.0000176)(6.1000000,0.0000194)(6.2000000,0.0000215)(6.3000000,0.0000237)(6.4000000,0.0000262)(6.5000000,0.0000290)(6.6000000,0.0000320)(6.7000000,0.0000354)(6.80000000,0.0000391)(6.9000000,0.0000432)(7.0000000,0.0000478)(7.1000000,0.0000528)(7.2000000,0.0000583)(7.3000000,0.0000645)(7.4000000,0.0000712)(7.5000000,0.0000787)(7.6000000,0.0000870)(7.7000000,0.0000962)(7.8000000,0.0001063)(7.9000000,0.0001175)(8.0000000,0.0001298)(8.1000000,0.0001435)(8.2000000,0.0001586)(8.3000000,0.0001752)(8.4000000,0.0001937)(8.5000000,0.0002140)(8.6000000,0.0002365)(8.70000000,0.0002614)(8.8000000,0.0002889)(8.9000000,0.0003193)(9.0000000,0.0003529)(9.1000000,0.0003900)(9.2000000,0.0004310)(9.3000000,0.0004763)(9.4000000,0.0005264)(9.5000000,0.0005818)(9.6000000,0.0006430)(9.7000000,0.0007106)(9.8000000,0.0007853)(9.9000000,0.0008679)(10.0000000,0.0009592)(10.1000000,0.0010601)(10.2000000,0.0011715)(10.3000000,0.0012947)(10.4000000,0.0014309)(10.50000000,0.0015814)(10.6000000,0.0017477)(10.7000000,0.0019314)(10.8000000,0.0021345)(10.9000000,0.0023590)(11.0000000,0.0026070)(11.1000000,0.0028812)(11.2000000,0.0031841)(11.3000000,0.0035189)(11.4000000,0.0038889)(11.5000000,0.0042977)(11.60000000,0.0047495)(11.7000000,0.0052489)(11.8000000,0.0058006)(11.9000000,0.0064104)(12.0000000,0.0070842)(12.1000000,0.0078288)(12.2000000,0.0086516)(12.3000000,0.0095609)(12.4000000,0.0105656)(12.5000000,0.0116758)(12.6000000,0.0129025)(12.7000000,0.0142579)(12.8000000,0.0157556)(12.9000000,0.0174104)(13.0000000,0.0192388)(13.1000000,0.0212588)(13.2000000,0.0234906)(13.3000000,0.0259561)(13.4000000,0.0286799)(13.5000000,0.0316888)(13.6000000,0.0350125)(13.7000000,0.0386838)(13.8000000,0.0427387)(13.9000000,0.0472172)(14.0000000,0.0521630)(14.1000000,0.0576245)(14.2000000,0.0636550)(14.30000000,0.0703131)(14.4000000,0.0776634)(14.5000000,0.0857769)(14.6000000,0.0947317)(14.7000000,0.1046137)(14.8000000,0.1155171)(14.90000000,0.1275454)(15.0000000,0.1408122)(15.1000000,0.1554419)(15.2000000,0.1715707)(15.3000000,0.1893476)(15.4000000,0.2089356)(15.5000000,0.2305121)(15.6000000,0.2542709)(15.70000000,0.2804224)(15.8000000,0.3091953)(15.9000000,0.3408374)(16.0000000,0.3756164)(16.1000000,0.4138210)(16.2000000,0.4557617)(16.3000000,0.5017709)(16.4000000,0.5522033)(16.5000000,0.6074355)(16.6000000,0.6678653)(16.7000000,0.7339105)(16.8000000,0.8060061)(16.9000000,0.8846014)(17.0000000,0.9701552)(17.1000000,1.0631300)(17.2000000,1.16398377)(17.3000000,1.2731600)(17.4000000,1.3910751)(17.5000000,1.5181031)(17.6000000,1.6545569)(17.7000000,1.8006653)(17.8000000,1.9565473)(17.9000000,2.12218058)(18.0000000,2.2973664)(18.1000000,2.4816895)(18.2000000,2.6744737)(18.3000000,2.8747338)(18.4000000,3.08112472)(18.5000000,3.2918888)(18.6000000,3.5048049)(18.7000000,3.7171407)(18.8000000,3.9256134)(18.9000000,4.12636244)(19.0000000,4.3149408)(19.1000000,4.4863307)(19.2000000,4.6349912)(19.3000000,4.7549431)(19.4000000,4.8398984)(19.5000000,4.8834379)(19.6000000,4.8792378)(19.7000000,4.8213437)(19.8000000,4.7044827)(19.9000000,4.52440192)(20.0000000,4.2782114)(20.10000000,3.9647082)(20.2000000,3.5846521)(20.30000000,3.1409632)(20.40000000,2.6388141)(20.5000000,2.0855955)(20.6000000,1.49074477)(20.7000000,0.8654382)(20.8000000,0.2221611)(20.9000000,-0.4258154)(21.0000000,-1.0650156)(21.1000000,-1.6823537)(21.20000000,-2.2656986)(21.3000000,-2.8043674)(21.4000000,-3.2895153)(21.5000000,-3.7143998)(21.6000000,-4.0745105)(21.7000000,-4.3675655)(21.8000000,-4.59338985)(21.90000000,-4.7536977)(22.0000000,-4.8518038)(22.1000000,-4.8922951)(22.2000000,-4.8806881)(22.3000000,-4.8230964)(22.4000000,-4.72592598)(22.50000000,-4.5956122)(22.6000000,-4.4384037)(22.7000000,-4.2601959)(22.8000000,-4.0664129)(22.9000000,-3.8619321)(23.0000000,-3.65104586)(23.1000000,-3.4374538)(23.20000000,-3.2242777)(23.3000000,-3.01409346)(23.4000000,-2.8089750)(23.5000000,-2.6105443)(23.6000000,-2.4200230)(23.7000000,-2.2382874)(23.8000000,-2.0659193)(23.9000000,-1.9032516)(24.0000000,-1.7504164)(24.1000000,-1.6073769)(24.2000000,-1.4739689)(24.30000000,-1.3499252)(24.4000000,-1.2349076)(24.5000000,-1.1285200)(24.6000000,-1.0303308)(24.70000000,-0.9398810)(24.8000000,-0.8567182)(24.9000000,-0.7803633)(25.0000000,-0.7103746)(25.1000000,-0.6462810)(25.2000000,-0.5876765)(25.3000000,-0.5341336)(25.4000000,-0.4852491)(25.5000000,-0.4406827)(25.6000000,-0.4000711)(25.7000000,-0.3631052)(25.80000000,-0.3294260)(25.9000000,-0.2987699)(26.0000000,-0.2709797)(26.1000000,-0.2457334)(26.2000000,-0.2226199)(26.3000000,-0.2018671)(26.4000000,-0.1828663)(26.5000000,-0.1657953)(26.6000000,-0.1499998)(26.7000000,-0.1361869)(26.8000000,-0.1232974)(26.9000000,-0.1116994)(27.0000000,-0.1010760)(27.1000000,-0.0914813)(27.2000000,-0.0825469)(27.3000000,-0.0744131)(27.4000000,-0.0663792)(27.5000000,-0.0609880)(27.6000000,-0.0549309)(27.7000000,-0.0510299)(27.8000000,-0.0454928)(27.9000000,-0.0429757)(28.0000000,-0.0379086)(28.1000000,-0.0407329)(28.2000000,-0.0353349)(28.30000000,-0.0265903)(28.4000000,-0.0202873)(28.5000000,-0.0348029)(28.60000000,-0.0314342)(28.7000000,-0.0220090)(28.8000000,-0.0119918)(28.9000000,-0.0122294)(29.0000000,-0.0225930)(29.1000000,0.0069359)(29.20000000,-0.0124471)(29.3000000,-0.0184181)(29.4000000,-0.0131379)(29.5000000,-0.0789299)(29.6000000,-0.0174004)(29.70000000,-0.0504483)(29.8000000,0.1183230)(29.9000000,0.0622612)(30.0000000,-0.0660134)
};
\addplot[color=black, no markers,dotted] coordinates{(-2,0)(32,0)};
\addplot[color=red, no markers,style={thick},dotted] coordinates{
(0.*10^-8,0.*10^-11)(0.1000000,0.*10^-11)(0.2000000,0.*10^-11)(0.3000000,0.*10^-11)(0.4000000,0.*10^-11)(0.5000000,0.*10^-11)(0.6000000,0.*10^-11)(0.7000000,0.*10^-11)(0.8000000,0.*10^-14)(0.9000000,0.*10^-11)(1.0000000,0.*10^-15)(1.1000000,0.*10^-11)(1.2000000,0.*10^-11)(1.3000000,0.*10^-14)(1.4000000,0.*10^-11)(1.5000000,0.*10^-15)(1.6000000,0.*10^-11)(1.7000000,0.*10^-11)(1.8000000,0.*10^-11)(1.9000000,0.*10^-15)(2.0000000,0.*10^-11)(2.1000000,0.*10^-11)(2.2000000,0.*10^-11)(2.3000000,0.*10^-11)(2.4000000,0.*10^-11)(2.5000000,0.*10^-11)(2.6000000,0.*10^-11)(2.7000000,0.*10^-11)(2.8000000,0.*10^-11)(2.9000000,0.*10^-10)(3.0000000,0.*10^-11)(3.1000000,0.*10^-11)(3.2000000,0.*10^-11)(3.3000000,0.*10^-11)(3.4000000,0.*10^-11)(3.5000000,0.*10^-11)(3.60000000,0.*10^-11)(3.7000000,0.*10^-11)(3.8000000,0.*10^-11)(3.9000000,0.*10^-11)(4.00000000,0.*10^-11)(4.1000000,0.*10^-13)(4.2000000,0.*10^-11)(4.3000000,0.*10^-11)(4.4000000,0.*10^-13)(4.5000000,0.*10^-11)(4.6000000,0.*10^-11)(4.70000000,0.*10^-11)(4.8000000,0.*10^-11)(4.9000000,0.*10^-11)(5.0000000,0.*10^-11)(5.1000000,0.*10^-11)(5.2000000,0.*10^-11)(5.3000000,0.*10^-11)(5.4000000,0.*10^-11)(5.5000000,0.*10^-11)(5.6000000,0.*10^-11)(5.7000000,0.*10^-11)(5.8000000,0.*10^-11)(5.9000000,0.*10^-11)(6.0000000,0.*10^-11)(6.1000000,0.*10^-11)(6.2000000,0.*10^-11)(6.3000000,0.*10^-11)(6.4000000,0.*10^-11)(6.5000000,0.*10^-11)(6.6000000,0.*10^-11)(6.7000000,0.*10^-11)(6.80000000,0.*10^-11)(6.9000000,0.*10^-12)(7.0000000,0.*10^-11)(7.1000000,0.*10^-11)(7.2000000,0.*10^-10)(7.3000000,0.*10^-10)(7.4000000,0.*10^-10)(7.5000000,0.*10^-10)(7.6000000,0.*10^-10)(7.7000000,0.*10^-10)(7.8000000,0.*10^-10)(7.9000000,0.*10^-10)(8.0000000,0.*10^-10)(8.1000000,0.*10^-10)(8.2000000,0.*10^-10)(8.3000000,0.*10^-10)(8.4000000,0.*10^-10)(8.5000000,0.*10^-9)(8.6000000,0.*10^-9)(8.70000000,0.*10^-9)(8.8000000,0.*10^-9)(8.9000000,0.*10^-9)(9.0000000,0.*10^-9)(9.1000000,0.*10^-9)(9.2000000,0.*10^-9)(9.3000000,0.*10^-9)(9.4000000,0.*10^-9)(9.5000000,0.*10^-9)(9.6000000,0.*10^-8)(9.7000000,0.*10^-8)(9.8000000,0.*10^-8)(9.9000000,0.*10^-8)(10.0000000,0.*10^-8)(10.1000000,0.*10^-8)(10.2000000,0.*10^-8)(10.3000000,0.*10^-8)(10.4000000,0.*10^-8)(10.50000000,0.*10^-8)(10.6000000,0.*10^-8)(10.7000000,1.*10^-7)(10.8000000,1.*10^-7)(10.9000000,1.*10^-7)(11.0000000,2.*10^-7)(11.1000000,2.*10^-7)(11.2000000,3.*10^-7)(11.3000000,3.*10^-7)(11.4000000,4.*10^-7)(11.5000000,5.*10^-7)(11.60000000,6.*10^-7)(11.7000000,7.*10^-7)(11.8000000,9.*10^-7)(11.9000000,1.1*10^-6)(12.0000000,1.3*10^-6)(12.1000000,1.6*10^-6)(12.2000000,2.0*10^-6)(12.3000000,2.5*10^-6)(12.4000000,3.0*10^-6)(12.5000000,3.7*10^-6)(12.6000000,4.5*10^-6)(12.7000000,5.5*10^-6)(12.8000000,6.7*10^-6)(12.9000000,8.1*10^-6)(13.0000000,9.9*10^-6)(13.1000000,0.0000121)(13.2000000,0.0000148)(13.3000000,0.0000181)(13.4000000,0.0000221)(13.5000000,0.0000270)(13.6000000,0.0000329)(13.7000000,0.0000402)(13.8000000,0.0000490)(13.9000000,0.0000599)(14.0000000,0.0000731)(14.1000000,0.0000892)(14.2000000,0.0001088)(14.30000000,0.0001328)(14.4000000,0.0001621)(14.5000000,0.0001977)(14.6000000,0.0002412)(14.7000000,0.0002942)(14.8000000,0.0003588)(14.90000000,0.0004374)(15.0000000,0.0005333)(15.1000000,0.0006500)(15.2000000,0.0007921)(15.3000000,0.0009651)(15.4000000,0.0011754)(15.5000000,0.0014312)(15.6000000,0.0017421)(15.70000000,0.0021198)(15.8000000,0.0025784)(15.9000000,0.0031347)(16.0000000,0.0038092)(16.1000000,0.0046263)(16.2000000,0.0056152)(16.3000000,0.0068111)(16.4000000,0.0082556)(16.5000000,0.0099982)(16.6000000,0.0120976)(16.7000000,0.0146233)(16.8000000,0.0176565)(16.9000000,0.0212924)(17.0000000,0.0256419)(17.1000000,0.0308326)(17.2000000,0.0370113)(17.3000000,0.0443442)(17.4000000,0.0530183)(17.5000000,0.0632406)(17.6000000,0.0752364)(17.7000000,0.0892459)(17.8000000,0.1055176)(17.9000000,0.1242979)(18.0000000,0.1458162)(18.1000000,0.1702634)(18.2000000,0.1977623)(18.3000000,0.2283284)(18.4000000,0.2618188)(18.5000000,0.2978672)(18.6000000,0.3358039)(18.7000000,0.3745595)(18.8000000,0.4125522)(18.9000000,0.4475612)(19.0000000,0.4765896)(19.1000000,0.4957245)(19.2000000,0.5000057)(19.3000000,0.4833195)(19.4000000,0.4383355)(19.5000000,0.3565124)(19.6000000,0.2282003)(19.7000000,0.0428640)(19.8000000,-0.2105491)(19.9000000,-0.5430822)(20.0000000,-0.9650883)(20.10000000,-1.4854126)(20.2000000,-2.11048815)(20.30000000,-2.84341279)(20.40000000,-3.6831002)(20.5000000,-4.6236065)(20.6000000,-5.65374057)(20.7000000,-6.7570397)(20.8000000,-7.9121627)(20.9000000,-9.0937103)(21.0000000,-10.2734221)(21.1000000,-11.4216593)(21.20000000,-12.5090267)(21.3000000,-13.5079968)(21.4000000,-14.3943696)(21.5000000,-15.1484426)(21.6000000,-15.7558138)(21.7000000,-16.2077658)(21.8000000,-16.5012468)(21.90000000,-16.6384979)(22.0000000,-16.6263914)(22.1000000,-16.4756088)(22.2000000,-16.1996864)(22.3000000,-15.8140965)(22.4000000,-15.33535798)(22.50000000,-14.78023527)(22.6000000,-14.1651337)(22.7000000,-13.5055430)(22.8000000,-12.8157445)(22.9000000,-12.10849865)(23.0000000,-11.3950129)(23.1000000,-10.6848292)(23.20000000,-9.9858601)(23.3000000,-9.30463947)(23.4000000,-8.6462115)(23.5000000,-8.0144285)(23.6000000,-7.41203149)(23.7000000,-6.8408457)(23.8000000,-6.3019242)(23.9000000,-5.7956818)(24.0000000,-5.3218604)(24.1000000,-4.8800499)(24.2000000,-4.4692887)(24.30000000,-4.0883606)(24.4000000,-3.7361755)(24.5000000,-3.4110884)(24.6000000,-3.1115413)(24.70000000,-2.8360832)(24.8000000,-2.5833884)(24.9000000,-2.3516439)(25.0000000,-2.1394842)(25.1000000,-1.9451144)(25.2000000,-1.7679214)(25.3000000,-1.6058786)(25.4000000,-1.4587598)(25.5000000,-1.3244387)(25.6000000,-1.2021435)(25.7000000,-1.0903306)(25.80000000,-0.9891279)(25.9000000,-0.8969045)(26.0000000,-0.8137822)(26.1000000,-0.7374330)(26.2000000,-0.6683190)(26.3000000,-0.6050428)(26.4000000,-0.5486324)(26.5000000,-0.4964250)(26.6000000,-0.4502225)(26.7000000,-0.4077546)(26.8000000,-0.3697218)(26.9000000,-0.3349912)(27.0000000,-0.3045559)(27.1000000,-0.2751856)(27.2000000,-0.2472023)(27.3000000,-0.2260451)(27.4000000,-0.2019643)(27.5000000,-0.1867007)(27.6000000,-0.1675548)(27.7000000,-0.1509888)(27.8000000,-0.1350169)(27.9000000,-0.1250671)(28.0000000,-0.1098772)(28.1000000,-0.0980693)(28.2000000,-0.0932103)(28.30000000,-0.0846766)(28.4000000,-0.0700817)(28.5000000,-0.0704961)(28.60000000,-0.0550994)(28.7000000,-0.0615418)(28.8000000,-0.0564595)(28.9000000,-0.0403230)(29.0000000,-0.0500996)(29.1000000,-0.0490151)(29.20000000,-0.0367805)(29.3000000,-0.0409316)(29.4000000,-0.0316823)(29.5000000,-0.0209903)(29.6000000,-0.0173576)(29.70000000,0.0020084)(29.8000000,-0.0323361)(29.9000000,-0.0402196)(30.0000000,-0.0319293)
};
\end{axis}
\node[] at (4.6,3.6) {b)};
\end{tikzpicture}
\caption{(a) \emph{Left}: reaction scheme for substrate inhibition. \emph{Right}: the stochastic production of P from S can be seen as a biased random walk from the state E to $E' \equiv E$ (identified by periodic conditions) through ES, with ESS being a trapping state from which escaping is only possible by unlikely fluctuations. (b) Mean reaction current (solid) given by \eqref{JSubIn} for the synthesis of dopamine. Kinetic parameters are in accord with physiological values \cite{reed10}: $k_2 [E]_{\textrm{tot}} =36 \,\mu\mathrm{M \,s}^{-1}$, $k_\mathrm{M}=46\,\mu\mathrm{M}$, $k_{-3}/k_3=160\,\mu\mathrm{M} $. The corresponding curve for Michaelis-Menten kinetics (dashed), i.e. $k_3=0$, plateaus only at large affinities. The shaded area indicates the range of daily affinities.  \emph{Inset}: the differential response $R$ given by \eqref{R} (solid) and the correlation $\mean{-\bar J_1( \bar W_{1}+  \bar W_{3})}_{\mathrm{cc}}$ responsible for NDR (dotted).}
\label{fig:SubIn1}
\end{figure}

Substrate inhibition is estimated to occur in 20\% of known enzymes~\cite{reed10}.
In its simplest form [see fig.~\ref{fig:SubIn1} (a)], it happens when up to two substrate molecules $S$ can bind the active site of one enzyme E giving an inert species ESS. The binding of a single substrate molecule results in the formation of the active complex ES decaying into the product P, as in the usual Michaelis-Menten scheme. 
The latter pathway is responsible for the production of P from S at a concentration rate $\mean{\bar J_1}$, that is the chemical current of biological interest. The former instead represents the competing process~\cite{hal30,reed10}. It takes up---or traps, within the mechanical analogy---substrate into ESS thus decreasing the rate of production of P for large [S] (fig.~\ref{fig:SubIn1}). Indeed, with [S]$\,\gg\,$[P] kept constant by particle reservoirs to mimic physiological conditions and fixing the reaction affinity $\mathcal{A}=\log \frac{k_1k_2 [S]}{k_{-1} k_{-2} [P]}$, the stationary current takes the non-monotonic form~\cite{hal30,reed10}
\begin{align}\label{JSubIn}
\mean{\bar J_1}= \frac{k_{2} \mathrm{[E]}_{\mathrm{tot}} \mathrm{[S]}  }{K_\textrm{M}+  \mathrm{[S]} +\frac{k_{3}}{k_{-3}} \mathrm{[S]}^2}\,.
\end{align}
Here $\mathrm{[E]}_{\mathrm{tot}}$ is the total concentration of enzyme and $K_\textrm{M}:=\frac{k_{2}  +k_{-1}}{k_{1}}$.
The kinetics of the usual Michaelis-Menten scheme is retrieved setting $k_3=0$ (fig.~\ref{fig:SubIn1}).

\begin{figure}[t]
\hspace{-0.4cm}
\centering

\caption{(a) Mean reaction current (solid) given by \eqref{JSubIn} and its scaled variance (dashed) for the synthesis of serotonin. Kinetic parameters are in accord with physiological values \cite{reed10}: $k_2 [E]_{\textrm{tot}} =36 \,\mu\mathrm{M \,s}^{-1}$, $k_\mathrm{M}=46\,\mu\mathrm{M}$, $k_{-3}/k_3=400\,\mu\mathrm{M} $. The shadowed area indicates the daily range of affinities.  \emph{Inset}: the differential response $R$ (solid) and the correlation $\mean{-\bar J_1( \bar W_{1}+  \bar W_{3})}_{\mathrm{cc}}$ responsible for NDR (dotted). (b) The signal-to-noise ratio $SNR$ (solid), and the upper bounds $\sqrt{\Sigma/2}$ (dashed) and $\sqrt{\sum_\rho \mean{\bar C_\rho}}$ (dotted) set by the uncertainty relations. The shadowed area indicates the range of daily affinities. The shaded area indicates the range of daily affinities.\emph{Inset}: parametric plot of $SNR$ for the two values of the affinity,   $A_{\mathrm{min}}$ and $A_{\mathrm{max}}$, corresponding to the same average current $\mean{\bar J_1}$.}
\label{fig:SubIn2}
\end{figure}
\begin{figure*}[t]
\hspace{0.2cm}
\includegraphics[width=0.20\textwidth]{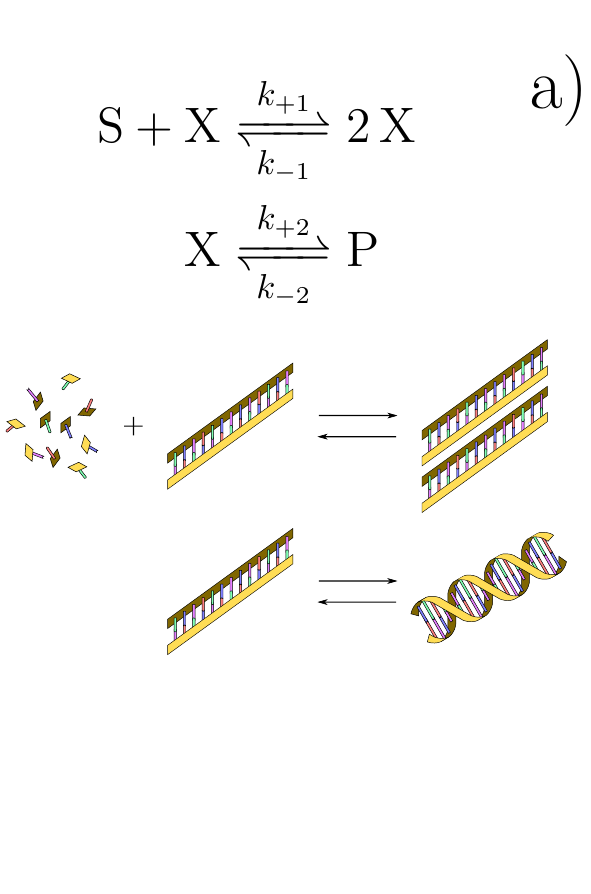}
\hspace{0.2cm}

\caption{(a) Minimal autocatalytic reaction scheme representing, e.g., a coarse-grained model of DNA replication. (b) Mean reaction current (solid) and its variance (dashed). Inset: the differential response $R$ (solid) and the correlation $\mean{\bar J_1 \bar F_{1}}_{\mathrm{cc}}$ responsible for NDR (dashed). (c) The signal-to-noise $SNR$ compared to the bounds $\sqrt{\Sigma/2}$ (dashed) and $ \sqrt{\sum_\rho \mean{\bar C_\rho}}$ (dotted) set by the uncertainty relations.}
\label{fig:AutoCat1}
\end{figure*}

The first two scaled cumulants of the time-averaged current  $\bar J_1$ show the existence of a marginal affinity $\mathcal{A}^*$ that marks the transition to a NDR regime, i.e. $R<0$ for $\mathcal{A}>\mathcal{A}^*$, where fluctuations $\textrm{Var}\bar J_1:= \mean{\bar J_1^2}_{\mathrm{cc}}$ peak.
In the present model of substrate inhibition, $-\mean{\bar J_1 (\bar W_1+\bar W_{3})}_{\mathrm{cc}}$ is the leading negative contribution in \eqref{R} for $\mathcal{A} \simeq \mathcal{A}^*$ (fig. \ref{fig:SubIn1}), confirming that ESS is a trapping state.

The existence of NDR has some crucial consequences. First,  since $R(\mathcal{A}^*)=0$, $\mean{\bar J_1}$ varies little upon sizable variations of substrate concentration around $[S](\mathcal{A}^*)$. Second, since $\mean{\bar J_1}$ is not an injective function of $\mathcal{A}$, a target mean current---e.g. required for optimal physiological functioning---is attainable at two different affinities $\mathcal{A}_{\mathrm{min}}$ and $\mathcal{A}_{\mathrm{max}}$. These two facts may constitute a crucial advantage to control environmental and intrinsic noise in biochemical systems. 

In the first case, the system can reach a homeostatic state characterized by a relative stable output $\mean{\bar J_1}$ despite variations in the environmental conditions, i.e. the substrate concentration [S]. Importantly, a similar stable regime would be achieved only at larger affinities in the absence of NDR, i.e. for the standard Michaelis-Menten kinetics (cf. fig.~\ref{fig:SubIn1}).
A representative example is the synthesis in neurons of dopamine (P) from tyrosine (S) mediated by the enzyme tyrosine hydroxylase (E)~\cite{nak99}. The tyrosine concentration in humans varies in response to meals on a timescale $\tau_{\textrm{S}} \sim 10^3\mathrm{s}$, and typically ranges from $100 \, \mu\textrm{M}$  to $120 \, \mu\textrm{M}$. Since the dynamics \eqref{cme} for the substrate inhibition scheme in fig.~\ref{fig:SubIn1} has a unique steady state, its typical relaxation timescale is well estimated by the inverse of the smallest (pseudo first-order) reaction rate constant, i.e. $(k_{-2}\mathrm{[P]})^{-1}=: \tau_{\textrm{rel}} \sim 1\mathrm{s}$. Hence, on the slow timescale $\tau_{\textrm{S}}$ the current $\bar J_1$ evolves quasi-statically, with all its moments depending parametrically on the instantaneous value of the affinity $21.0  \lesssim \mathcal{A} \lesssim 21.2 $ (in units of $RT=1$). This interval is placed very close to $\mathcal{A}^* \simeq 20.8$, hence resulting in a current relative variation smaller than 3\%.

In the second case, the system can increase the (scaled) signal-to-noise ratio $SNR:= \mean{\bar J_1}/\sqrt{\textrm{Var}{\bar J_1}}$ selecting the optimal affinity among $\mathcal{A}_{\mathrm{min}}$ and $\mathcal{A}_{\mathrm{max}}$. Consider, for example, the synthesis of serotonin (P) out of tryptophan (S) catalyzed by tryptophan hydoxylase (E) in human cells~\cite{fri72,mck05}. For different values of the parameters compatible with physiological conditions, we found that $SNR$ is always smaller at $\mathcal{A}_{\mathrm{min}}$, i.e. higher precision is achieved at $\mathcal{A}<\mathcal{A}^*$ (fig.~\ref{fig:SubIn2}).
As a consequence, such robustness against intrinsic fluctuations is achieved at the smaller value of the mean dissipation rate $\Sigma(\mathcal{A}):= \mathcal{A} \mean{\bar J_1}$. Remarkably, the daily concentration of tryptophan $ 25 \, \mu\textrm{M} \lesssim \mathcal{A} \lesssim  35 \, \mu\textrm{M}$ \cite{reed10} yields a range of affinities $19.6  \lesssim \mathcal{A} \lesssim 19.9 $ which  is close to optimal in order to maximize $SNR$. 
 Thanks to stochastic uncertainty relations \cite{bar15a,pol16,hor17,pro17,dec18}, $SNR$ can be bounded by dissipation, $SNR \leqslant \sqrt{\Sigma/2}$, and by the system's dynamical activity, $ SNR \leqslant \sqrt{\sum_\rho \mean{\bar C_\rho}}$. The entropic bound means that a more precise current \emph{may} be obtained at larger affinity, and thus dissipation. Nevertheless, such condition need not be realized in practice, especially because the bound becomes looser as $\mathcal{A}$ increases, as is the case for serotonin synthesis.

\section{Autocatalysis} 

Autocatalysis represents the second scenario in which NDR can arise, whose simplest possible scheme is depicted in fig.~\ref{fig:AutoCat1} (a). Having one dynamical concentration, two reactions (required to have a maximum current), and two fixed concentrations [S] and [P] (needed to set the system away from equilibrium), this is the minimal chemical scheme displaying NDR. An outstanding example falling into the autocatalytic paradigm is DNA replication~\cite{pla11}: two double stranded molecules are produced by one such molecule (X) and nucleobases (S), and eventually undergo a conformational change, e.g. into the double helix structure (P). Several other biological processes can be similarly described at a coarse-grained level as autocatalytic reactions, e.g. formation of micelles from amphiphiles~\cite{col18,bac92}, ATP net production in glycolysis~\cite{che11}, and conversion of prion proteins into the infectious form~\cite{bie04}.
Here, we regard the autocatalytic scheme as a model for phosphorylation of protein kinase (X) coupled to a larger association/dissociation cycle via the conversion into the complex P~\cite{wang02,meh06,pla11}.  For the chosen physiological values of the parameters, the coupled cycle is known to display circadian rhythmicity~\cite{meh06}. Taking [P] as time-independent, we highlight the role played by NDR in triggering chemical oscillations, a topic of major relavance which may even have a role in our understanding of the origin of life~\cite{sem16}.
We consider the degradation of S into P as the current of interest. Its macroscopic value determined by the rate equation,
\begin{align}\nonumber
\mean{\bar J_1}=\frac {k_ 2} {2 k_ {-1}}  \bigg( &\sqrt {4 k_ {-2} k_ {-1} [P] + \left (k_ 2 - k_ 1 [S] \right)^2} \\
& + k_ 1 [S] -k_ 2 \bigg) - k_ {-2} [P]\, ,
\end{align}
is a non-monotonic function of [P], and so of the affinity $\mathcal{A}=\log \frac{k_1 k_2  [S]}{k_{-1} k_{-2} [P]}$.
Moreover, the full statistics of the model can be obtained by the large deviation techniques introduced in section 2. The negative correlation $\mean{\bar J_1 \bar F_{1}}_{\mathrm{cc}}<0$ [see inset in fig.~\ref{fig:AutoCat1} (b)] entering \eqref{R} shows that NDR is induced by a competition between forward and backward flows (due to the nonlinearity of the autocatalytic step), rather than by the presence of a trapping state. Also, the qualitative behavior of the $\text{Var} \bar J_1$ is different, with the minimum (rather then the maximum) occurring near $\mathcal{A}^*$.
Despite that, for a wide range of parameters compatible with physiological conditions we observe that $SNR$ is larger for $\mathcal{A}<\mathcal{A}^*$, i.e. at smaller dissipation.  Hence, as already discussed for substrate inhibition, autocatalysis can be run at low affinities to reduce the current dispersion or around the region of null response to minimize variations in the output current. 

\section{Dissipative self-assembly}

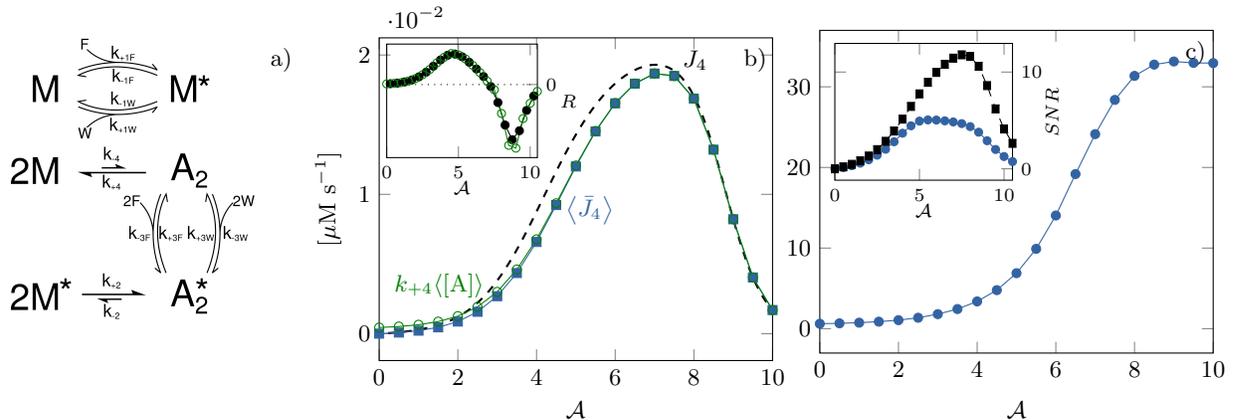
\begin{figure*}[t]
\hspace{0.4cm}
\includegraphics[width=0.18\textwidth]{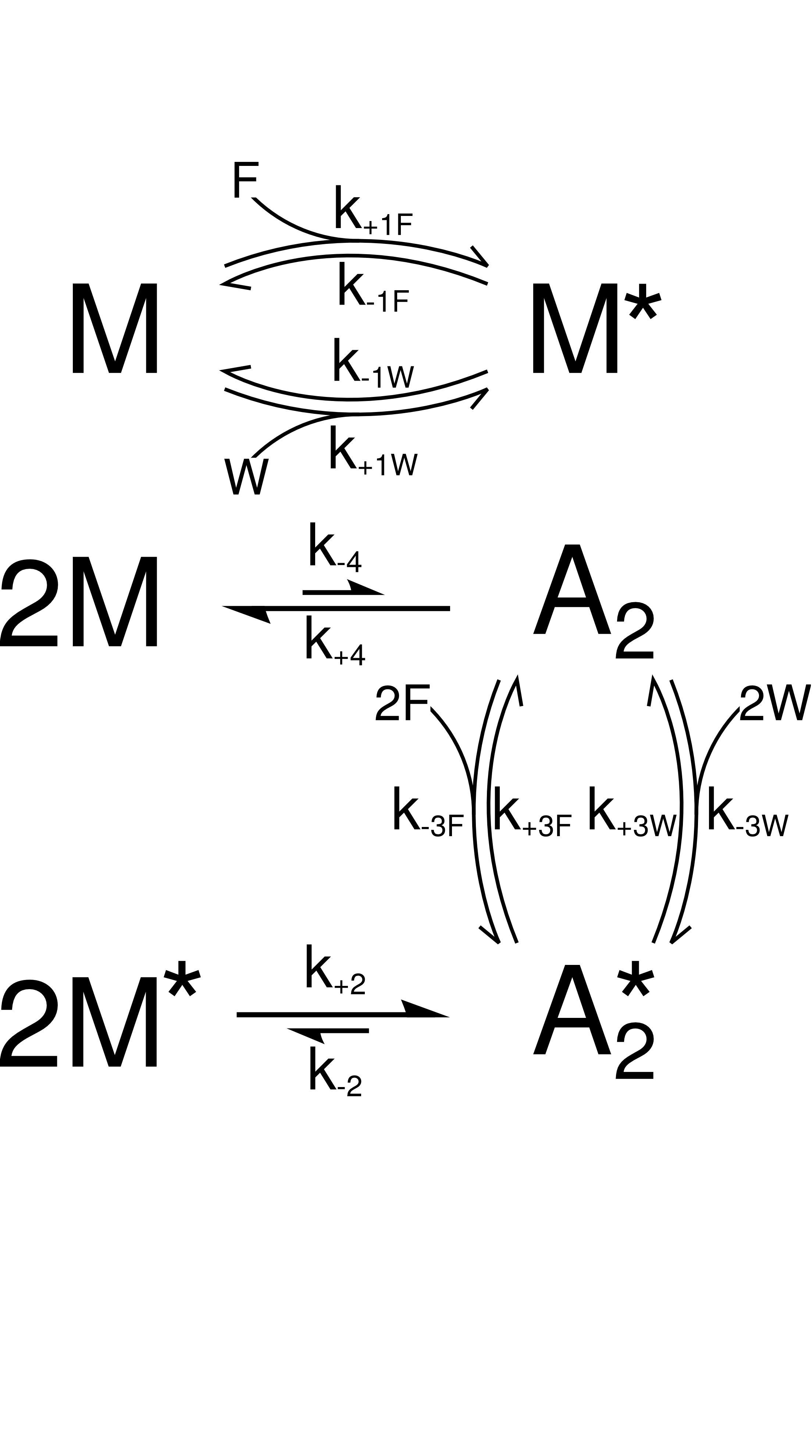}
\begin{tikzpicture}
\node[] at (-1,4) {a)};
\hspace{0.3cm}
\begin{axis}[width=0.38\textwidth,xmin=-0,xmax=10, xlabel=$\mathcal{A}$, ylabel= {[$\mu$M s$^{-1}$]},
 xlabel near ticks, ylabel near ticks] 
\addplot[color=black, no markers,style={thick},dashed] coordinates{
( -0.176430397605521 , -2.326786129622574e-05 )
( -0.06713970786400399 , -9.418742858390765e-06 )
( 0.042150981877514544 , 6.300492323366418e-06 )
( 0.15144167161903307 , 2.4160109822402216e-05 )
( 0.2607323613605501 , 4.447093466571409e-05 )
( 0.3700230511020671 , 6.759011241436988e-05 )
( 0.4793137408435857 , 9.392746625303829e-05 )
( 0.5886044305851043 , 0.00012395242720689354 )
( 0.6978951203266213 , 0.00015820147969863635 )
( 0.8071858100681397 , 0.00019728601951409864 )
( 0.9164764998096583 , 0.0002419004648006617 )
( 1.0257671895511753 , 0.0002928303923547678 )
( 1.1350578792926924 , 0.0003509603912155185 )
( 1.2443485690342109 , 0.0004172812344719187 )
( 1.3536392587757293 , 0.0004928958704644376 )
( 1.462929948517248 , 0.000579023630039702 )
( 1.5722206382587665 , 0.0006770019429095078 )
( 1.6815113280002836 , 0.0007882847613593107 )
( 1.790802017741802 , 0.0009144368138311353 )
( 1.9000927074833205 , 0.001057122767036085 )
( 2.0093833972248376 , 0.00121809037833462 )
( 2.1186740869663563 , 0.0013991467870907454 )
( 2.227964776707873 , 0.001602127242279486 )
( 2.3372554664493923 , 0.0018288558105970833 )
( 2.44654615619091 , 0.0020810979679442097 )
( 2.555836845932428 , 0.0023605054540977747 )
( 2.6651275356739466 , 0.002668554359790261 )
( 2.7744182254154635 , 0.0030064780999181683 )
( 2.8837089151569812 , 0.0033751976631401833 )
( 2.9929996048985 , 0.0037752522601176572 )
( 3.1022902946400177 , 0.004206734137310902 )
( 3.2115809843815355 , 0.0046692317865664485 )
( 3.320871674123054 , 0.005161785963361126 )
( 3.430162363864572 , 0.005682862740576902 )
( 3.53945305360609 , 0.006230347211517674 )
( 3.648743743347608 , 0.006801560404361272 )
( 3.7580344330891258 , 0.007393300529801505 )
( 3.8673251228306444 , 0.008001907968366634 )
( 3.9766158125721613 , 0.008623351584136232 )
( 4.085906502313679 , 0.009253332232465807 )
( 4.195197192055198 , 0.009887397919281352 )
( 4.304487881796716 , 0.010521064144829752 )
( 4.413778571538235 , 0.011149932636763126 )
( 4.523069261279753 , 0.011769801979462095 )
( 4.63235995102127 , 0.012376764508792736 )
( 4.7416506407627885 , 0.012967285148283027 )
( 4.850941330504306 , 0.013538259415986582 )
( 4.960232020245824 , 0.014087049447408902 )
( 5.069522709987342 , 0.014611498374560432 )
( 5.17881339972886 , 0.015109924637323989 )
( 5.288104089470378 , 0.015581098695721832 )
( 5.397394779211896 , 0.01602420513036841 )
( 5.506685468953415 , 0.016438793280800633 )
( 5.615976158694932 , 0.016824719429065035 )
( 5.7252668484364495 , 0.01718208315975633 )
( 5.834557538177968 , 0.01751115999427074 )
( 5.943848227919487 , 0.01781233177873099 )
( 6.053138917661004 , 0.018086015664163853 )
( 6.162429607402522 , 0.018332591904746013 )
( 6.271720297144041 , 0.018552330156435613 )
( 6.381010986885558 , 0.018745313520813368 )
( 6.490301676627076 , 0.018911359287374762 )
( 6.599592366368594 , 0.019049935234561263 )
( 6.708883056110112 , 0.01916007053254386 )
( 6.818173745851629 , 0.01924026086335839 )
( 6.927464435593149 , 0.019288368500393838 )
( 7.036755125334666 , 0.019301519991933146 )
( 7.146045815076184 , 0.019276007050556734 )
( 7.255336504817703 , 0.01920720056537091 )
( 7.36462719455922 , 0.019089493580340462 )
( 7.473917884300737 , 0.018916296644631368 )
( 7.5832085740422555 , 0.018680117671394963 )
( 7.692499263783774 , 0.01837276690959934 )
( 7.801789953525291 , 0.01798573294799244 )
( 7.9110806432668115 , 0.01751077304451893 )
( 8.020371333008327 , 0.016940743999952712 )
( 8.129662022749846 , 0.016270661546855955 )
( 8.238952712491363 , 0.01549891345224122 )
( 8.348243402232882 , 0.014628469866856827 )
( 8.457534091974399 , 0.013667854039617254 )
( 8.566824781715917 , 0.012631592563482612 )
( 8.676115471457436 , 0.011539896879358184 )
( 8.785406161198953 , 0.010417459526794667 )
( 8.894696850940473 , 0.009291460540322816 )
( 9.00398754068199 , 0.008189103663436816 )
( 9.113278230423509 , 0.007135147442321979 )
( 9.222568920165028 , 0.006149897060864301 )
( 9.331859609906545 , 0.005247978821574954 )
( 9.441150299648061 , 0.004437995869307959 )
( 9.550440989389578 , 0.003722952138832919 )
( 9.659731679131097 , 0.003101199336886018 )
( 9.769022368872616 , 0.0025676282677578113 )
( 9.878313058614133 , 0.002114868599517497 )
( 9.987603748355653 , 0.0017343405222257358 )
( 10.09689443809717 , 0.0014170826800906644 )
( 10.206185127838689 , 0.0011543431938120796 )
( 10.315475817580207 , 0.0009379588379582054 )
( 10.424766507321724 , 0.000760564734590019 )
( 10.534057197063243 , 0.000615679872406738 )
( 10.643347886804762 , 0.0004977087775390465 )
};
\addplot[color=skyblue2, mark=cube*,mark size=1.7] coordinates{
( 0.0 , 0.0 )
( 0.5 , 0.0774519555556 * 10^-3 )
( 1.0 , 0.2134012777778 * 10^-3)
( 1.5 , 0.4524993777778 * 10^-3)
( 2.0 , 0.8682791333333 * 10^-3)
( 2.5 , 1.5691394111111 * 10^-3)
( 3.0 , 2.6898769666667 * 10^-3)
( 3.5 , 4.3518184222222 * 10^-3)
( 4.0 , 6.5812439555556 * 10^-3)
( 4.5 , 9.2340459777778 * 10^-3)
( 5.0 , 11.9965909222222 * 10^-3)
( 5.5 , 14.5141685555556 * 10^-3)
( 6.0 , 16.5332471111111 * 10^-3)
( 6.5 , 17.9438794222222 * 10^-3)
( 7.0 , 18.6716531333333 * 10^-3)
( 7.5 , 18.4962214444444 * 10^-3)
( 8.0 , 16.8692753777778 * 10^-3)
( 8.5 , 13.2040296111111 * 10^-3)
( 9.0 , 8.2117916222222 * 10^-3)
( 9.5 , 4.0339233666667 * 10^-3)
( 10.0 , 1.6947110555556 * 10^-3)
( 10.5 , 0.6581788444444 * 10^-3)
};
\addplot[color=webgreen, mark=o,mark size=1.7] coordinates{
( 0.0 , 0.43940383370739966* 10^-3  )
( 0.5 , 0.5113892427909397 * 10^-3 )
( 1.0 , 0.6384138871425296 * 10^-3 )
( 1.5 , 0.8632847538729993 * 10^-3 )
( 2.0 , 1.2572235622370191* 10^-3  )
( 2.5 , 1.9254213180525184* 10^-3  )
( 3.0 , 3.0001199505832177 * 10^-3 )
( 3.5 , 4.603538447834667 * 10^-3 )
( 4.0 , 6.7665719251882654 * 10^-3 )
( 4.5 , 9.355552210329463* 10^-3  )
( 5.0 , 12.06772240731833 * 10^-3 )
( 5.5 , 14.550810457325147* 10^-3  )
( 6.0 , 16.550989878301365 * 10^-3 )
( 6.5 , 17.951173991979747 * 10^-3 )
( 7.0 , 18.674662047231056* 10^-3  )
( 7.5 , 18.497429574288553* 10^-3  )
( 8.0 , 16.86945297265072 * 10^-3 )
( 8.5 , 13.20409121912028 * 10^-3 )
( 9.0 , 8.211978297695264 * 10^-3 )
( 9.5 , 4.033864925460296* 10^-3  )
( 10.0 , 1.6949465705945288 * 10^-3 )
( 10.5 , 0.6583767033966396 * 10^-3 )
};
 \end{axis}
  \node[] at (5,4) {b)};
 \begin{axis}[anchor=south west, xshift=0.1cm, yshift=2.7cm ,xmin=0, xmax=10.5,width=0.2\textwidth, ylabel=$R$,
 xlabel=$\mathcal{A}$, 
 style={font=\scriptsize},  ylabel style={rotate=0},  ytick={0}, ylabel near ticks, yticklabel pos=right,  xlabel near ticks, xlabel shift= -5 pt,  ylabel shift = -5pt, ylabel style={rotate=-90}]
\addplot[color=black,mark size=1.5,mark=*] coordinates{
( 0.25 , 0.1549816 )
( 0.75 , 0.2718986444444 )
( 1.25 , 0.47819620000000007 )
( 1.75 , 0.8315595111109999 )
( 2.25 , 1.4017205555555998 )
( 2.75 , 2.2414751111112 )
( 3.25 , 3.323882911111 )
( 3.75 , 4.458851066666799 )
( 4.25 , 5.305604044444401 )
( 4.75 , 5.5250898888888 )
( 5.25 , 5.0351552666667985 )
( 5.75 , 4.038157111110998 )
( 6.25 , 2.821264622222202 )
( 6.75 , 1.4555474222222031 )
( 7.25 , -0.35086337777779875 )
( 7.75 , -3.2538921333332027 )
( 8.25 , -7.330491533333401 )
( 8.75 , -9.984475977777798 )
( 9.25 , -8.355736511111001 )
( 9.75 , -4.678424622222201 )
( 10.25 , -2.0730644222223997 )
};
\addplot[color=black, no markers,dotted] coordinates{(-2,0)(10,0)};
\addplot[color=webgreen, mark=o, mark size=1.5,solid] coordinates{
( 0.0 , 0.11491683272928033 )
( 0.5 , 0.20396057617695937 )
( 1.0 , 0.3549684448095471 )
( 1.5 , 0.6192854446455758 )
( 2.0 , 1.0854565238475349 )
( 2.5 , 1.787194193358291 )
( 3.0 , 2.735822065002765 )
( 3.5 , 3.9175982243759075 )
( 4.0 , 4.935418889375542 )
( 4.5 , 5.613823387785572 )
( 5.0 , 5.3667475334237995 )
( 5.5 , 4.623514169781208 )
( 6.0 , 3.4740566194398594 )
( 6.5 , 2.166016671414118 )
( 7.0 , 0.6256892802284426 )
( 7.5 , -0.9439684067801863 )
( 8.0 , -5.228087500370009 )
( 8.5 , -11.003345757853594 )
( 9.0 , -11.513269381662383 )
( 9.5 , -6.1583581925110025 )
( 10.0 , -3.3834769210614013 )
( 10.5 , -1.2334335480363343 )
};
\end{axis}
\node[] at (0.8,1) {\textcolor{webgreen}{$k_{+4}\mean{\textrm{[A]}}$}};
  \node[] at (2.8,2) {\textcolor{skyblue2}{$\mean{\bar J_4}$}};
    \node[] at (4.2,4.) {$J_4$};
 \end{tikzpicture}
 \begin{tikzpicture}
\begin{axis}[width=0.38\textwidth,xmin=0, xmax=10, ymax=37, xlabel=$\mathcal{A}$,  ylabel style={rotate=0}]
\addplot[color=skyblue2,mark size=1.7,mark=*] coordinates{
( 0.0 , 0.6147951035884408 )
( 0.5 , 0.6641195804226739 )
( 1.0 , 0.7425814452323145 )
( 1.5 , 0.8705814205271898 )
( 2.0 , 1.0566889766334056 )
( 2.5 , 1.3554145706482352 )
( 3.0 , 1.803410386440505 )
( 3.5 , 2.4353140580580392 )
( 4.0 , 3.3809331662173765 )
( 4.5 , 4.782026607413468 )
( 5.0 , 6.883312533919717 )
( 5.5 , 9.900737999243228 )
( 6.0 , 14.038395652825596 )
( 6.5 , 19.179042585975488 )
( 7.0 , 24.12870621995112 )
( 7.5 , 28.374140736317802 )
( 8.0 , 31.35420165857751 )
( 8.5 , 32.75381003950691 )
( 9.0 , 33.14219387614048 )
( 9.5 , 32.96586564515178 )
( 10.0 , 32.942210357508934 )
( 10.5 , 32.876710218286235 )
};
\end{axis}
 \begin{axis}[anchor=south west, xshift=0.2cm, yshift=2.3cm ,xmin=0, xmax=10.5,width=0.22\textwidth, ylabel=$SNR$,
 xlabel=$\mathcal{A}$, 
 style={font=\scriptsize},  ylabel style={rotate=0},  ytick={0,10}, ylabel near ticks, yticklabel pos=right, xlabel near ticks,  xlabel shift= -5 pt, ylabel shift = -5pt]
 \addplot[color=skyblue2, mark=*, mark size=1.5] coordinates{
( 0.0 , 0.0 )
( 0.5 , 0.1278448468072734 )
( 1.0 , 0.316019158121452 )
( 1.5 , 0.5777970078420341 )
( 2.0 , 0.9285965336076107 )
( 2.5 , 1.4029114690864486 )
( 3.0 , 2.018557866980041 )
( 3.5 , 2.794041850237232 )
( 4.0 , 3.678743701186481 )
( 4.5 , 4.444687004088275 )
( 5.0 , 4.926683650470906 )
( 5.5 , 5.050789480764741 )
( 6.0 , 5.035712933609453 )
( 6.5 , 4.953922191814898 )
( 7.0 , 4.84117515887692 )
( 7.5 , 4.670223250423986 )
( 8.0 , 4.360878559160494 )
( 8.5 , 3.782088934705929 )
( 9.0 , 2.899737439562689 )
( 9.5 , 1.9956209002861054 )
( 10.0 , 1.2675684933903641 )
( 10.5 , 0.7438298186236246 )
};
\addplot[color=black, mark size=1.5, mark=cube*,dashed] coordinates{
( 0.0 , 0.0 )
( 0.5 , 0.19678917088549358 )
( 1.0 , 0.4619537615149378 )
( 1.5 , 0.8238622862267091 )
( 2.0 , 1.3177853644150856 )
( 2.5 , 1.9806182185817007 )
( 3.0 , 2.8407095768487314 )
( 3.5 , 3.9027380744520506 )
( 4.0 , 5.130787056799609 )
( 4.5 , 6.446177696899156 )
( 5.0 , 7.7448663391378805 )
( 5.5 , 8.934647561910642 )
( 6.0 , 9.959893707598821 )
( 6.5 , 10.799778527564548 )
( 7.0 , 11.432478818407366 )
( 7.5 , 11.778015997328795 )
( 8.0 , 11.616979083316902 )
( 8.5 , 10.594066815649425 )
( 9.0 , 8.59686713867324 )
( 9.5 , 6.190498524620909 )
( 10.0 , 4.116686842055879 )
( 10.5 , 2.62885485842528 )
};
\end{axis}
 \node[] at (5,4) {c)};
 \end{tikzpicture}
\caption{(a) General scheme of dissipative self-assembly. (b) The reaction current given by the rate equations (solid), its mean as obtained from the stochastic simulations at finite number of chemicals  (filled) and its approximation \eqref{currconc} (open). \emph{Inset}: the differential response $R$ obtained from \eqref{R} by stochastic simulations (filled) and from numerical derivative of the mean current (open). (c) Signal-to-noise ratio of the concentration $[\mathrm{A}_2]$ obtained from stochastic simulations. The bound $\sqrt{\Sigma/2}$ does not apply. \emph{Inset}: $SNR$ of the current $\mean{\bar J_4}$ (circle) compared to the bound offered by the square root of half the mean dissipation rate $\sqrt{\Sigma/2}$ (square), both obtained from stochastic simulations. The bound  $ \sqrt{\sum_\rho \mean{\bar C_\rho}}$ is not shown, being one order of magnitude larger.}
\label{fig:DissSelfAss}
\end{figure*}

As a final example, we analyze dissipative self-assembly, a paradigm of out-of-equilibrium synthesis extensively exploited by biological systems: prominent examples being the formation of microtubules out of tubulin dimers fueld by guanosine 5'-triphosphate (GTP)~\cite{des97,hes17} and the ATP-driven self-assembly of actin filaments~\cite{how01}. It has been also 
 probed in experiments such as the controlled gelation of dibenzoyl-L-cysteine to form nanofibers~\cite{boe10} and the chemically fueled transient self-assembly of fibrous hydrogel materials~\cite{boe15}.
A simple, yet insightful model is sketched in fig.~\ref{fig:DissSelfAss} (a), which has been proposed as a minimal general scheme for genuine nonequilibrium self-assembly~\cite{rag18}.
The direct aggregation of two monomers (M) to form the assembled state ($\mathrm{A}_2$)---which would be highly disfavored at equilibrium---is boosted by coupling the process with the burning reaction of some fuel (F) converted into waste (W). This fueling mechanism opens side pathways involving the activated species $\mathrm{M}^*$, which easily aggregates into $\mathrm{A}_2^*$.
To give an example, supposing M to not aggregate because of unfavorable electrostatic interactions, then F (W) may be a high (low) energy methylating agent able to convert negatively charged monomers M into their neutral form $\mathrm{M}^*$.
By properly fixing the concentrations of F and W, a nonequilibrium stationary state rich in the target species $\mathrm{A}_2$ can be achieved.
At odds with conventional equilibrium self-assembly, the efficacy of this synthetic procedure is not determined by the relative thermodynamic stabilities of the components, but rather by the sustained dissipation and kinetic aspects~\cite{otto15,sor17,ros17}.

By design, the system attains large concentrations of $\mathrm{A}_2$ depleting the monomer concentration $[\mathrm{M}]$~\cite{rag18}.
Therefore, the current of reaction $\rho=4$---which is half the current from F to W---is almost unidirectional, especially far from equilibrium:
 \begin{align}\label{currconc}
 \mean{\bar J_4} = k_{+4} [\mathrm{A}_2]\, .
 \end{align}
Because of the proportionality relation~\eqref{currconc}, the existence of NDR affecting  $\mean{\bar J_4}$ sets an upper bound on the maximal $[\mathrm{A}_2]$  achievable by the process.

Being unable to calculate \eqref{scgf} for this model, we performed stochastic simulations based on the Gillespie algorithm \cite{gil77}. We  measured the mean current and its variance, as well as its response, for different values of the affinity $\mathcal{A}=\log \frac{k_{+1F}^2k_{+2}k_{+3W}k_{+4}[F]^2}{k_{-1F}^2k_{-2}k_{-3}k_{-4}[W]^2}$.
The response was obtained by directly measuring $ \mean{\bar J_4}$ at different $\mathcal{A}$, and through~\eqref{R}, by estimating the required correlation functions. The good agreement of the two methods (fig. \ref{fig:DissSelfAss}) shows that~\eqref{R} is not only conceptually revealing, but also of practical relevance for calculating responses without actually applying perturbations. Despite their proportionality in average, the current $\bar J_4$ and the concentration $[\mathrm{A}_2]$ were found to possess different fluctuations. It implies that the signal-to-noise ratio $\mean{[\text{A}_2]}/\sqrt{\text{Var}[\text{A}_2]}$ is not bounded by dissipation, hence does not decrease at large $\mathcal{A}$ due to NDR. 
Indeed, $\mean{[\text{A}_2]}/\sqrt{\text{Var}[\text{A}_2]}$ is close to its maximum at the optimal affinity $\mathcal{A}^*$. This is important for the scalability of artificial syntesis to microscopic volumes.

\section{Discussion}

In conclusion, we have shown that NDR is a widespread phenomenon in chemistry with major consequences on the efficacy of biological and artificial processes. In substrate inhibition, NDR allows a system to reach homeostasis at lower dissipation than in the Michaelis-Menten kinetics, keeping the signal-to-noise ratio unaltered. For systems that do not need to maintain a stable current,  
higher precision to sustain a given mean current can be reached at low affinity, i.e. dissipation. 

Since the analogous behavior was found in both biochemical schemes, despite the difference in the qualitative behavior of the current fluctuations, the idea that life efficiency always increases with the dissipation rate is called into question \cite{bai18}. 
Still, it is worth noticing that whenever these chemical schemes are used as effective models that coarse-grain some \emph{nonequilibrium} reactions, the dissipation $\Sigma$ is always smaller than the total entropy production rate of the original process \cite{esp12}. Instead, if only \emph{equilibrated} subprocesses are lumped or discarded, a complete thermodynamic description of the original process exists \cite{wac18}. It identifies $\mathcal{A}$ with the chemical potential difference of the fixed species (respectively, P and S, F and W) and $\Sigma$ with the entropy production rate \cite{rao16}.
Remarkably---given the scarcity of solvable models away from equilibrium---these results, obtained in the large-size limit, are exact. They show that the general bounds offered by the uncertainty relations have little predictive power for parameters that are biologically relevant. 

Lastly, we have shown with stochastic simulations that NDR limits the efficacy of  
 dissipative self-assembly: the ideal affinity that maximizes the output mean concentration also yields a nearly optimal signal-to-noise ratio. 
Altogether, we have achieved a fundamental analysis of NDR of reaction currents. It pinpoints the relation between robustness, precision and dissipation in biochemistry, and allows the optimization of performance and scalability in nonequilibrium synthesis.  

\section{Acknowledgements}
We thank Arthur Watchel for help with the numerical simulations, which were carried out using the HPC facilities of the University of Luxembourg~\cite{VBCG_HPCS14}{\small -- see \url{https://hpc.uni.lu}}.

\begin{widetext}

\appendix

\section{Stochastic dynamics of chemical reaction networks}
Consider a well-mixed volume $V$ occupied by dilute reacting chemical species $X_\sigma$, labelled by the index $\sigma \in \{1,\dots,M\}$, following mass action kinetics. The population number of the dynamical species $\vec n=(n_1,\dots, n_M)$ varies in time because of the random reactions, while the concentration of the externally controlled species, $c_{\sigma'}:=[X_{\sigma'}]$ with $\sigma'=\{M+1, \dots N\}$, is kept constant. A single reactive event, occurring thorough the reaction $\rho \in \{\pm 1, \dots, \pm \mathcal{M}  \}$, involves $\nu_{\sigma, \rho}$ molecules $\sigma$ and changes the population of species $\sigma$ as $n_\sigma \longrightarrow n_\sigma + S_{\sigma, \rho}$, with $S_{\sigma, \rho}:=\nu_{\sigma, -\rho} -\nu_{\sigma, +\rho} $ the stoichiometric 
coefficient.
For compactness, we will denote $\vec S_\rho$ the vector of the stoichiometric coefficients corresponding to reaction $\rho$. The  reactions happen with a probability rate 
\begin{align}\label{WVa}
W^{(V)}_{\rho}(\vec n) = V k_{\rho} \prod_{\sigma'=M+1}^Nc_{\sigma'}^{\nu_{\sigma', \rho}}  \prod_{\sigma=1}^M \frac{1}{V^{\nu_{\sigma, \rho}}} \frac{n_\sigma !}{ (n_\sigma - \nu_{\sigma, \rho}) !}
\end{align}
with $k_\rho$ being the rate constant. 
The stochastic dynamics can be described by the chemical master equation
\begin{align}\label{cmea}
\partial_t P_t(\vec n)&= \sum_{\rho=-\mathcal{M}}^{\mathcal{M}}  \left[W^{(V)}_\rho(\vec n-\vec S_\rho )P_t(\vec n-\vec S_\rho)-W^{(V)}_\rho(\vec n )P_t(\vec n)\right]\\
&=\sum_{\rho=-\mathcal{M}}^{\mathcal{M}}  \left[ \exp(-\vec S_\rho \cdot \partial_{\vec n}) -1 \right] W^{(V)}_\rho(\vec n)P_t(\vec n) \nonumber \\
& =: H^{(V)}(\vec n, -\partial_{\vec n}) P_t(\vec n),
\end{align}
that prescribes the time evolution of the probability $P_t(\vec n)$ of the chemical populations in terms of the action of the operator $H^{(V)}(\vec n, \partial_{\vec n})$.
The solution of \eqref{cmea} can be used to study only the statistics of state-like observables, i.e. functions of the instantaneous population $\vec n$. In order to obtain the statistics of transition-like observables it is convenient to resort to a path integral representation of the probability of full stochastic trajectories. For example, the probability $\mathcal{P}[\vec n(t)]$ of the population trajectory $\{\vec n(t): t \in (0,T] \}$ can be obtained from \eqref{cmea} introducing auxiliary variables $\vec p$---to be marginalized over, eventually---that accounts for variations in population $\partial_{\vec n}$: 
\begin{align}\label{Ppatha}
\mathcal{P}[\vec n(t)] = \int \mathcal D \vec p \, e^{\int_0^T \mathrm{d}t \left[- \dot {\vec n}(t) \cdot \vec p(t) + H^{(V)}(\vec n(t), \vec p(t)) \right]}
\end{align}
Two observables are of interest to us, namely, the time-averaged  number of reactive events $\rho$, $\bar C^{(V,T)}_\rho := \frac{1}{T} \int_0^T \mathrm{d} t \, \delta_{\rho, \rho(t)}$, and the time-averaged reaction rate $\bar W^{(V,T)}[\vec n(t)] := \frac 1 T \int_0^T \mathrm{d} t \, W^{(V)}(\vec n(t))$.
Within this formalism, the full statistics of the above observables is encoded in the cumulant generating function
\begin{align}\label{cgfa}
g^{(V,T)}(q, \lambda)&:=\log \mean{e^{T\sum_{\rho} (q_\rho \bar C^{(V,T)}_\rho + \lambda_\rho \bar W^{(V,T)}_\rho)}}\\
&=  \int \mathcal D \vec n  \, \int \mathcal D \vec p \,e^{\int_0^T \mathrm{d}t \left[- \dot {\vec n}(t) \cdot \vec p(t) + H^{(V)}_{q ,\lambda}(\vec n(t), \vec p(t)) \right]}\nonumber
\end{align}
computed by functional integration of a path probability with `tilted' generator
 \begin{align}\label{Hqa}
&H^{(V)}_{q ,\lambda}(\vec n, \vec p) := \sum_{\rho=-\mathcal{M}}^{\mathcal{M}}  \left[ \exp(\vec S_\rho \cdot \vec p +  q_\rho) +\lambda_\rho-1 \right] W^{(V)}_\rho(\vec n).
 \end{align}
The superscript $(V,T)$ stands for the dependence on a finite system volume $V$ and trajectory duration $T$. Later, we will omit the superscripts $V$ and $T$ to indicate the large $V$ and $T$ limit of the various functions.
All cumulants, such as the mean $\mean{\bar C^{(V,T)}_\rho}=\partial_{q_\rho} g^{(V,T)}(q,\lambda)|_{q,\lambda=0}$ and the connected correlations, e.g, 
\begin{align}
\partial_{q{_\rho}}\partial_{q{_{\rho'}}}g^{(V,T)}(q,\lambda)|_{q,\lambda=0} = \langle \bar C^{(V,T)}_{\rho} \bar C^{(V,T)}_{\rho'}  \rangle_{\mathrm{cc}}:=\langle \bar C^{(V,T)}_{\rho} \bar C^{(V,T)}_{\rho'} \rangle- \langle \bar C^{(V,T)}_{\rho} \rangle \langle \bar C^{(V,T)}_{\rho'}  \rangle,
\end{align}
can be calculated from \eqref{cgfa} upon differentiation. The statistics of the time-integrated current, $\bar J^{(V,T)}_\rho:=(\bar C^{(V,T)}_{+\rho}- \bar C^{(V,T)}_{-\rho})$, follows from \eqref{cgfa} as well. 

\subsection{Macroscopic limit: the rate equations}

In the thermodynamic limit, given by $n_{\sigma} \to \infty$, $V \to \infty$ and finite concentrations $c_\sigma:= n_\sigma /V= [X_\sigma]$, the probability $P_t(\vec n)$ becomes sharply peaked around its maximum. Thereby, one obtains the chemical rate equations multiplying \eqref{cmea} by $\vec n$ and averaging,
\begin{align}\label{rea}
&
\dot{c}_\sigma =\sum_{\rho=-\mathcal{M}}^{\mathcal{M}}  S_{\sigma, \rho} W_{\rho}(\vec c) 
\end{align}
with the average reaction current given by the (scaled) large-size limit of \eqref{Wa}:
\begin{align}\label{Wa}
W_{\rho}(\vec c) := k_{\rho} \prod_{\sigma'=M+1}^N{c_{\sigma'}}^{\nu_{\sigma', \rho}}  \prod_{\sigma=1}^M {c_\sigma}^{\nu_{\sigma, \rho}}.
\end{align}
The same result can be obtained by the path integral formalism. The statistical weight in \eqref{Ppatha} peaks as $ \exp V \int_0^T \mathrm{d}t [- \dot {\vec c}(t) \cdot \vec p(t) + H(\vec c(t), \vec p(t)) ]$
around those paths that maximize the time integral, i.e.,those that satisfy the Hamilton equations
\begin{align}
&\dot {\vec p}= -\partial_{\vec c } H(\vec c, \vec p) && \dot {\vec c}= \partial_{\vec p } H(\vec c, \vec p),
\end{align}
with $H(\vec c(t), \vec p(t))$ being now function of the macroscopic rates \eqref{Wa}. The rate equations \eqref{rea} are regain by looking for the noise-less trajectories:
\begin{align}
&\vec p= \vec 0\quad  \Longrightarrow \quad \dot {\vec c}= \partial_{\vec p } H(\vec c, \vec p)|_{\vec p= \vec 0}=\sum_\rho S_{\sigma, \rho} W_{\rho}(\vec c) .
\end{align}

\subsection{Macroscopic limit: scaled cumulant generating function}

Fluctuations in the thermodynamics limit are captured by the scaled cumulant generating function
\begin{align}\label{scgfa}
g(q, \lambda):=\lim_{\substack{V \to \infty \\ T\to \infty}} \frac{1}{V \,T}  g^{(V,T)}(q, \lambda)= \lim_{\substack{V \to \infty \\ T\to \infty}} \frac 1 V  \int \mathcal D \vec c  \, \int \mathcal D \vec p \,e^{V \int_0^T \mathrm{d}t \left[- \dot {\vec c}(t) \cdot \vec p(t) + H_{q ,\lambda}(\vec c(t), \vec p(t)) \right]}.
\end{align}
The average \eqref{scgfa} is dominated by the trajectories the maximize the statistical weight, namely, by the solutions of the Hamilton equations
\begin{align}\label{hama}
&\dot {\vec p}= -\partial_{\vec c } H_{q ,\lambda}(\vec c, \vec p) && \dot {\vec c}= \partial_{\vec p } H_{q ,\lambda}(\vec c, \vec p).
\end{align}
Since here we are interested only in systems with a single stable stationary state we focus on the unique fixed point of \eqref{hama}---we thus assume the absence of multiple stable fixed points and time-dependent attractors for \eqref{hama}, which excludes the emergence in the thermodynamic limit of ergodicity breaking and limit-cycles, respectively. Namely, we seek the vectors $\vec c^*$ and $\vec p^*$ solution of
\begin{align}\label{hamsta}
&\vec 0= -\partial_{\vec c } H_{q ,\lambda}(\vec c, \vec p) &&\vec 0= \partial_{\vec p } H_{q ,\lambda}(\vec c, \vec p).
\end{align}
To avoid clutter we do not explicitly write the parametric dependence of $\vec c^*$ and $\vec p^*$ on the counting fields $q_\rho$ and $\lambda_\rho$. 
If there exist vectors $\vec \ell^\lambda$ such that $\vec \ell^\lambda \cdot \vec S_\rho=0 \, \forall \rho$, then the dynamics \eqref{cmea} conserves the concentrations $\vec \ell_\lambda \cdot \vec c$. Therefore, \eqref{hama} needs to be supplemented by the constraints
\begin{align}
& \sum_{\sigma=1}^M c_\sigma \ell_\sigma^\lambda = \mathrm{const}, && \sum_{\sigma | \ell_\sigma \neq 0 } \frac{p_\sigma}{ \ell_\sigma^\lambda }=0 .
\end{align}
Using the solution $\vec c^*$ and $\vec p^*$ to evaluate \eqref{scgfa} yields the scaled cumulant generating function
\begin{align}\label{scgfaa}
g(q, \lambda)= H_{q ,\lambda}(\vec c^*, \vec p^*) .
\end{align}
By virtue of the above assumptions, \eqref{scgfaa} is a smooth function of $q_\rho$ and $\lambda_\rho$.

\section{Response of chemical currents}

Consider the perturbation $c_{\sigma'} \to c_{\sigma'}(1+\epsilon) $ in the concentration of a one fixed species $\sigma'$, with $\epsilon \ll 1$. We are then interested in the response of a time-integrated current $\bar J_\rho$, a long time after the application of the perturbation---for simplicity, we work in the large $T$ and $V$ limit, although these results can be equally derived for finite $T$ and $V$.
Such response is defined as 
\begin{align}\label{Rdefa}
R:= \frac{\partial \mean{\bar J_\rho}_\epsilon}{\partial \epsilon}\bigg|_{\epsilon=0}=\frac{\partial }{\partial \epsilon} ( \partial_{q_\rho}-\partial_{q_{-\rho}})  g_\epsilon(q,\lambda)  \bigg|_{\substack{q=0,\lambda=0\\ \epsilon=0}}  .
\end{align}
Here, the subscript $\epsilon$ indicates that the scaled cumulant generating function $\eqref{scgfaa}$ corresponds to the 
dynamics with perturbed reaction rates 
\begin{align}
W_{\tilde \rho}(\vec c) (1+ \nu_{\sigma', \tilde \rho} \epsilon) + O(\epsilon^2),
\end{align}
where $\tilde \rho$ labels the reactions whose rate depends explicitly on the perturbed species $\sigma'$.
Performing the derivatives in \eqref{Rdefa}, the response function is then expressed as
\begin{align}\label{Ra}
R&=\sum_{\tilde \rho} \nu_{\sigma',\tilde \rho}  \left[ \mean{\bar J_\rho  \bar  C_{\tilde \rho}  }_{\mathrm{cc}}   - \mean{\bar J_\rho  \bar  W_{\tilde \rho} }_{\mathrm{cc}}  \right]\nonumber\\
&=\sum_{\tilde \rho} \nu_{\sigma',\tilde \rho}  \left[ \frac 1 2 \mean{\bar J_\rho  \bar  J_{\tilde \rho}  }_{\mathrm{cc}} +
\frac 1 2 \mean{\bar J_\rho  \bar  F_{\tilde \rho}  }_{\mathrm{cc}}  - \mean{\bar J_\rho  \bar  W_{\tilde \rho} }_{\mathrm{cc}}  \right]
\end{align}
where in the last line we have added and subtracted the same quantity to obtain the (large $V$ and $T$ limit of) time-averaged current $\bar  J_{\tilde \rho}$ and the time average traffic $ \bar F_{\tilde \rho}:= \bar C_{\tilde \rho} + \bar C_{-\tilde \rho}$. The latter counts the number of times the reaction channel $\tilde \rho$ has been used, in both forward ($+\tilde \rho$) and backward ($-\tilde \rho$) direction. 
 Notice that \eqref{Ra} contains only unperturbed averages.

When the unperturbed state coincides with equilibrium, defined as $\mathcal A=0$, the second and third term in \eqref{Ra} vanish, since they are averages of time-antisymmetric observables done with a time-symmetric measure.
Therefore, \eqref{Ra} simplifies to
\begin{align}
 R
&=\sum_{\tilde \rho} \nu_{\sigma',\tilde \rho}    \mean{\bar J_\rho  \bar  J_{\tilde \rho}  }_{\mathrm{cc}} 
\end{align}
Perturbing an equilibrium state the response is only dissipative or, equivalently, the dynamical contributions coincides with the dissipative one. If $ \bar J_{\rho}$ and  $\bar  J_{\tilde \rho}$ are not independent cycle currents \cite{and07, rao16}, their covariance vanish identically, so that $R \propto   \mean{\bar  J_{\rho}^2  }_{\mathrm{cc}} >0$ holds true at equilibrium.

\section{Substrate Inhibition}

In the thermodynamic limit, the tilted generator of the substraste inhibition scheme reads 
\begin{equation}
\begin{aligned}
\label{HSubIna}
H_{q ,\lambda}(\vec c, \vec p)& = k_{1} c_{\sS} c_{\sE} \left(e^{-p_{\sE}+p_{\sES}+q_1} -1 + \lambda_1\right)
+ k_{-1} c_{\sES} \left(e^{p_{\sE}-p_{\sES}+q_{-1}} -1\right) \\
&\quad+ k_{2} c_{\sES} \left(e^{-p_{\sES}+p_{\sE}+q_2} -1\right)
+ k_{-2} c_{\sP} c_{\sE} \left(e^{-p_{\sE}+p_{\sES}+q_{-2}} -1\right) \\
&\quad+ k_{3} c_{\sS} c_{\sES} \left(e^{-p_{\sES}+p_{\sESS}+q_3}-1+ \lambda_3\right)
+  k_{-3} c_{\sESS} \left(e^{-p_{\sESS}+p_{\sES}+q_{-3}}-1\right),
\end{aligned}
\end{equation}
where the reactions are numbered as in fig.~1~(a) of the main text. Note that for the sake of clarity we have identified the labels $\sigma$ and $\sigma'$ with the species name. Also, in view of \eqref{Ra}, we have added the counting field $\lambda$ only on the reactions $+1$ and $+3$, whose rates depend explicitly on the perturbed species S.
The dynamics conserves the total concentration of enzyme $c_{\text{E}_\text{tot}}=c_\text{E}+c_\text{ES}+c_\text{ESS}$, so that $\vec \ell=(1,1,1)$.

The rate equations \eqref{rea} obtained from \eqref{HSubIna} by setting $\lambda=q=p_\sigma=0 \, \forall \sigma$ read 
\begin{align}
\dot  c_{\sE}&= - \underbrace{(k_{1}  c_{\sS}c_{\sE}  - k_{-1}  c_{\sES})}_{\mean{J_1}} +  \underbrace{k_{2} c_{\sES} - k_{-2} c_{\sP} c_{\sE}}_{ \mean{J_2}}    \\
\dot  c_{\sES}&= \underbrace{(k_{1}  c_{\sS}c_{\sE}  - k_{-1}  c_{\sES})}_{\mean{J_1}} -  \underbrace{k_{2} c_{\sES} - k_{-2} c_{\sP} c_{\sE}}_{ \mean{J_2}}  -  \underbrace{(k_{3}  c_{\sES} c_{\sS}  - k_{-3}  (c_{\text{E}_\text{tot}}- c_\text{E}-c_\text{ES}))}_{\mean{J_3}},
\end{align}
where we have eliminated $c_\text{ESS}=c_{\text{E}_\text{tot}}- c_\text{E}-c_\text{ES}$. 
The stationary conditions $\dot  c_{\sE}=0$ and $\dot  c_{\sES}=0$ imply that $\mean{J_3}=0$ and
\begin{align}
 \mean{J_1}=\mean{J_2} =  \frac{k_{2} c_{\text{E}_\text{tot}} \left ( c_\text{S} -  c_\text{P} \frac{k_{-1}k_{-2}}{k_{1}} \right) }{K_\textrm{M}+c_\text{P} \frac{k_{-2}}{k_{1}} +  c_\text{S} \left( 1 +c_\text{P} \frac{k_{-2}}{k_{1}} \right) + \frac{k_{3}}{k_{-3}}c_\text{S}^2} \underset{k_{-2}c_\text{P} \ll 1 } {\simeq}  \frac{k_{2} c_{\text{E}_\text{tot}}  c_\text{S}  }{K_\textrm{M}+  c_\text{S}  + \frac{k_{3}}{k_{-3}}c_\text{S}^2}
\end{align}
that is Eq. (1).

For $\lambda$, $q$ (and so $p_\sigma$) different from zero, the Hamilton equations \eqref{hamsta} for the concentrations  can be solved under the constraint  $c_\text{E}+c_\text{ES}+c_\text{ESS} = c_{\text{E}_\text{tot}}$ and $p_\text{E}+p_\text{ES}+p_\text{ESS}=0$,
obtaining 
\begin{equation}
\begin{aligned}\label{c(p)}
&c_\text{E}(\vec p)= c_{\text{E}_\text{tot}} e^{4 p_\sE+2p_\sES+q_{-3}}  \left(k_{-1} e^{q_{-1}}+k_2 e^{q_2}\right) f(\vec p) ,\\
&c_\text{ES}(\vec p)= c_{\text{E}_\text{tot}} e^{2 p_\sE +4 p_\sES+q_{-3}} \left(k_{-2} c_\sP e^{q_{-2}}+k_1 e^{q_1} c_\sS \right) f(\vec p),\\
& f(\vec p)= k_{-2} c_\sP e^{2 p_\sE+4 p_\sES+q_{-2}+q_{-3}}+ k_{-2} \frac{k_3}{k_{-3}} c_\sP s e^{\text{q2m}+\text{q3}}+ k_1 c_\sS e^{2 p_\sE+4 p_\sES+q_1+q_{-3}}\\
& \quad \qquad + k_{-1} e^{4 p_\sE+2p_\sES+q_{-1}+q_{3}}
+k_2 e^{4 \text{pe}+2 \text{pes}+\text{q2}+\text{q3m}}+k_1 \frac{k_3}{k_{-3}} c_\sS^2 e^{q_1+q_3}.
\end{aligned}
\end{equation}
The constraint Hamilton equations for $\vec p$ are most easily solved by the change of variables $\log \psi =p_\sE+2 p_\sES$ and $\log \phi =p_\sES-p_\sP$, that yields
\begin{align}
& \phi(\psi)  = \frac{k_{-3} \left(e^{q_{-3} }\psi -1\right) + k_{-2} c_\sP + k_1 c_\sS e^{q_{-1} } }{ k_{-2} c_\sP + k_1 c_\sS e^{q_{-1} }(1-\lambda_1)} 
\\&\psi [k_{-1} e^{q_{-2}} +k_2 e^{q_2} - (k_{-1}  + k_2) \phi(\psi) ]+\phi(\psi)[ k_3  c_\sS \left(e^{q_3} +\psi(\lambda_3-1)\right)-\psi k_{-3} \left(e^{q_{-3} }\psi -1\right)]
=0
\end{align}
The latter is a 3rd order ordinary differential equation with constant coefficients, whose solutions can be expressed in closed form. We avoid to report them here, being too lengthy. The only physical solution $\vec p^*$ is the one giving positive concentrations $\vec c^*$ when plugged into \eqref{c(p)}. Finally, the scaled cumulant generating function is obtained inserting $\vec c^*$ and $\vec p^*$ into \eqref{HSubIna}, according to \eqref{scgfaa}.

Concerning the numerical values of the rate constants, for both the examples in the main text --- i.e., tyrosine hydroxylase (TH) and tryptophan hydroxylase (TPH) --- we relied on experimentally available $K_\text{M} = \frac{k_{2} + k_{-1}}{k_{1}}$ and $K_\text{i} = \frac{k_{-3}}{k_{3}}$~\cite{reed10}.
Within these constrains, we chose realistic $k_\rho$'s based on literature typical values~\cite{pur10}.
In particular, we have set $k_{3} < k_{1}$, thus considering negative cooperativity between molecules $\mathrm{S}$ upon their binding to the enzyme $\mathrm{E}$.
$k_{-3}$ has been kept small to make the ``trapping effect" well highlighted, while $k_{-1}$ and $k_{2}$ have been chosen in order to make reaction 2 rate limiting. $k_{-2}$ is usually neglected in kinetic models, but here it guarantees thermodynamic consistency.
Since the two enzyme considered have same $K_\mathrm{M}$ and different $K_\mathrm{i}$, we have opted to keep differences minimal.
Accordingly to the above argumentation, we checked the robustness of the qualitative features shown by the model under different choices.
Plots in the main text were obtained with the following parameters:
\begin{table}[h!]
\begin{tabular}{l|ll}
  & TH & TPH \\ \hline
  $K_\text{M}$     & $46 \, \mu \text{M}$ & $46 \, \mu \text{M}$ \\
  $K_\text{i}$     & $160 \, \mu  \text{M}$ & $400 \, \mu  \text{M}$ \\
  $k_{1}$  & $1 \, \mu \text{M}^{-1} \text{s}^{-1}$ &  $1 \,\text{M}^{-1} \text{s}^{-1}$\\
  $k_{-1}$  &  $20 \,\text{s}^{-1}$ &  $20 \,\text{s}^{-1}$ \\
  $k_{2}$  & $26\, \text{s}^{-1}$&  $26 \text{s}^{-1}$ \\
  $k_{-2}$  & $0.1 \, \cdot 10^{-6} \, \mu\text{M}^{-1} \text{s}^{-1}$ & $0.025  \cdot 10^{-6} \, \mu\text{M}^{-1} \text{s}^{-1}$ \\
  $k_{3}$  & $3 \cdot 10^{-1} \, \mu\text{M}^{-1} \text{s}^{-1}$ & $10^{-1} \, \mu\text{M}^{-1} \text{s}^{-1}$ \\
  $k_{-3}$  &  $48\, \text{s}^{-1}$&  $10\,  \text{s}^{-1}$\\
  $[\text{E}]_\text{tot}$ &  $1 \mu \text{M}$ & $ 1 \mu \text{M}$ \\
  $[\text{P}]$ &  $1\,\mu \text{M}$ &  $1\, \mu \text{M}$ 
\end{tabular}
\end{table}

\section{Autocatalysis}
In the thermodynamic limit, the tilted generator of the autocatalytic scheme reads 
\begin{equation}
\begin{aligned}\label{Hautoa}
H_{q ,\lambda}(\vec c, \vec p)& = k_{1} c_{\sS} c_{\sX} \left(e^{p_{\sX}} -1 +\right)
+ k_{-1} c_{\sX}^2 \left(e^{-p_{\sX}} -1\right) \\
&\quad+ k_{2} c_{\sX} \left(e^{-p_{\sX}+q_2} -1\right)
+ k_{-2} c_{\sP} \left(e^{p_{\sX}+q_{-2}} -1 \right )
\end{aligned}
\end{equation}
where the reactions are numbered as in~fig.~2~(a). Note that for the sake of clarity we have identified the labels $\sigma$ and $\sigma'$ with the species name. Also, we do not need any counting field $\lambda$, since the rate $W_{-2}$ is a constant, hence it does not contribute to the last term in \eqref{Ra}.

 The rate equation \eqref{rea} obtained from \eqref{Hautoa} by setting $\lambda=q=p_\sigma=0 \, \forall \sigma$ read 
\begin{align}\label{reAutoa}
\dot c_\sX = \underbrace{k_{1} c_\sS c_\sX - k_{-1}  c_\sX^2}_{\mean{J_1}} -\underbrace{(k_{2}  c_\sX -  k_{1} c_\sP)}_{\mean{J_2}}
\end{align}
At stationarity $\dot c_\sX =0$ we find 
\begin{align}
c_\sX^\pm= \frac{ c_\sP  k_{1}- k_{2} \pm \sqrt{\left(c_\sS  k_{1}- k_{2} \right )^2 + 4 c_\sP k_{-1}k_{-2}} }{2k_{-1}} ,
\end{align}
where $c_\sX^-<0$ (for all choice of parameters) is discarded because unphysical.
The stationary current is then obtained using $c_\sX^+$,
\begin{align}
\mean{J_1}=\mean{J_2}= \frac {k_ 2} {2 k_ {-1}}  \bigg( &\sqrt {4 k_ {-2} k_ {-1} c_\sP + \left (k_ 2 - k_ 1 c_\sS \right)^2} + k_ 1 c_\sS -k_ 2 \bigg) - k_ {-2} c_\sP,
\end{align}
that has a maximum al long as $4 k_ {-2} k_ {-1} c_\sP + k_{1} c_\sS (k_{1} c_\sS-2  k_{2}) <0 $.

For $q$ and then $p_\sX$ different from zero, we first solved \eqref{hamsta} for $c_\sX$
\begin{align}\label{cxa}
c_\sX(p_\sX)= \frac{e^{p_\sX}(1-e^{p_\sX})c_\sS k_1 +k_2 (e^{q_2}-1)}{2k_{-1} (e^{p_\sX}-1)}.
\end{align}
The resulting 5th-order ordinary differential equation for $p_\sX$ was solved numerically, as it doesn't allow a general close-form expression, and then inserted back into \eqref{cxa}. The so obtained $c_\sX^*$ and $p_\sX^*$ gives the scaled cumulant generating function according to \eqref{scgfaa}.

Plots in the main text where obtained with the following parameters, directly taken from from Ref.~\cite{meh06} (Figure 2) by considering the species KaiAC* as the dynamical variable (X). 
\begin{table}[h!]
\begin{tabular}{l|l}
 $ k_{1}$ &  $2 \cdot 10^{-3} \, \mu \text{M}^{-1} \text{s}^{-1} $ \\  
 $k_{-1}$ &  $0.4 \cdot 10^{-4} \, \mu \text{M}^{-1} \text{s}^{-1} $ \\
 $ k_{2}$ &  $8 \cdot 10^{-3} \,  \text{s}^{-1} $ \\
 $k_{-2}$ &  $ 4 \cdot 10^{-4} \,  \text{s}^{-1} $\\
 $[\text{S}]$ & $ 4 \, \mu \text{M}$ 
\end{tabular}
\end{table}

\section{Dissipative self-assembly}

We have run the standard Gillespie algorithm considering a total population of 1000 molecules  M and generating 10$^5$ trajectories of duration 1000 time units.
In order to ensure stationarity, we have sampled the trajectories after a period 100 time units that was previously checked to be long enough for the relaxation of the chemical network for all values of affinities. This allowed us to calculate mean values and covariances. The macroscopic current $J_4$  plotted in fig. 3 (b) was obtained by numerical integration of the rate equations \eqref{rea}. Its values does not coincide with the average $\mean{ \bar J_4}$ since the latter pertains to a system with large, yet finite number of molecules.
All the plots in the main text were obtained with the following macroscopic parameters:
\begin{table}[h!]
\begin{tabular}{l|l}
 $k_\mathrm{+1F}$ & $5.00      \,        \text{M}^{-1} \text{s}^{-1}$ \\
 $k_\mathrm{-1F}$ & $2.24\cdot10^{-2}  \, \text{s}^{-1}$               \\
 $k_\mathrm{+1W}$ & $1.00\cdot10^{-3}   \, \text{M}^{-1} \text{s}^{-1}$ \\
 $k_\mathrm{-1W}$ & $3.75\cdot10^{-2}  \, \text{s}^{-1}$               \\
 $k_\mathrm{+2}$  & $1.00 \,             \text{M}^{-1} \text{s}^{-1}$ \\
 $k_\mathrm{-2}$  & $1.80\cdot10^{-1}  \, \text{s}^{-1}$               \\
 $k_\mathrm{+3F}$ & $1.00\cdot10^{-6} \, \text{s}^{-1}$               \\
 $k_\mathrm{-3F}$ & $5.82\cdot10^{+1}  \, \text{M}^{-2} \text{s}^{-1}$ \\
 $k_\mathrm{+3W}$ & $2.00\cdot10^{+1}  \, \text{s}^{-1}$               \\
 $k_\mathrm{-3W}$ & $1.66\cdot10^{+1}  \, \text{M}^{-2} \text{s}^{-1}$ \\
 $k_\mathrm{+4} $ & $1.00\cdot10^{-1}  \, \text{s}^{-1}$               \\
 $k_\mathrm{-4} $ & $4.79\cdot10^{-4}  \, \text{M}^{-1} \text{s}^{-1}$ \\
 $[\text{M}]_\text{tot}$ &  $1\, \text{M}$ \\
 $[\text{W}]$  &  $1\, \text{M}$ 
\end{tabular}
\end{table}

\end{widetext}

\end{document}